\documentclass[a4paper,twoside]{article}
\newcommand{\affil}[1]{$^{\rm #1}$}
\usepackage[authoryear]{natbib}
\bibpunct{(}{)}{;}{a}{}{,}
\usepackage{graphicx}
\usepackage{amsmath}
\usepackage{amssymb}
\usepackage{setspace}
\usepackage{multicol}
\usepackage{rotating}
\date{} 
\title{\large\bf\flushleft Polarisation of Class II Methanol Masers}
\author{\parbox{\textwidth}{\flushleft
\vspace{-0.5cm}
{\it P. D. Stack\affil{A}, S. P. Ellingsen\affil{A}}}\\
\vspace{0.4cm}
{\small \affil{A}\,School of Mathematics and Physics, University of Tasmania, Private Bag 37, Hobart, Tasmania 7001, Australia}\\}
\begin{document}
\twocolumn[
\begin{changemargin}{.8cm}{.5cm}
\begin{minipage}{.9\textwidth}
\vspace{-1cm}
\maketitle
%
%
\small{\bf Abstract:}

We have used the University of Tasmania Mt Pleasant 26m radio telescope to investigate the polarisation characteristics of a sample of strong 6.7 GHz methanol masers, the first spectral line polarisation observations to be undertaken with this instrument.  As part of this process we have developed a new technique for calibrating linear polarisation spectral line observations.  This calibration method gives results consistent with more traditional techniques, but requires much less observing time on the telescope.  We have made the first polarisation measurements of a number of 6.7 GHz methanol masers and  find linear polarisation at levels of a few - 10\% in most of the sources we observed, consistent with previous results.  We also investigated the circular polarisation produced by Zeeman splitting in the 6.7 GHz methanol maser G9.62+0.20 to get an estimate of the line of sight magnetic field strength of 35$\pm$7 mG.


\medskip
\medskip
\end{minipage}
\end{changemargin}
]
\small

\section{Introduction}

Interstellar masers from a variety of molecules, including OH, water and methanol have been detected in interstellar space.
The geometry of the methanol molecule gives it a very rich rotational spectrum with more than a thousand transitions at frequencies less than 1 THz \citep[see for example][]{Cragg05}. To date more than twenty of these transitions have been observed to exhibit maser emission towards high-mass star formation regions.  Although many of these transitions are relatively weak and rare, the 6.7 GHz $5_{1}\mbox{-}6_{0}\mbox{~A}^{+}$ transition is the second strongest interstellar maser transition (after the 22 GHz transition of water) and has been detected towards more than 900 regions within our Galaxy \citep[see][and references therein]{Caswell10,Green10}.  This transition is exclusively detected towards young, high-mass star formation regions \citep{Minier03,Xu08}.  The role of magnetic fields in regulating star formation is still a matter of active debate, measurements of the polarisation properties of interstellar masers have the potential to reveal details about the orientation and strength of the field at very high resolution for a few select lines of sight within star formation regions \citep[e.g.][]{Dodson08,Vlemmings10,Surcis11}.  Although a few objects have been studied in detail, to date there have been few studies which attempt to look at the polarisation properties of 6.7~GHz methanol masers in general \citep[the exceptions being][]{Vlemmings08,Vlemmings11}.  

Here we present the results of a preliminary study of both the linear and circular polarisation properties of 6.7 GHz methanol masers undertaken with the University of Tasmania 26m Mt Pleasant radio telescope.  The maximum percentage linear polarisation detected in previous 6.7 GHz methanol maser studies ranges from a few to 10\% \citep{Ellingsen02,Dodson08}, while the fraction of circular polarisation is even smaller ($<$ 0.5\%) \citep{Vlemmings08}.  Even though 6.7~GHz methanol maser emission can be very strong (peak flux densities in excess of 1000 Jy), the majority of the emission in the majority of sources is significantly weaker.  Accurate measurement of the polarisation properties of the masers hence requires careful calibration of the instrument.  The observations undertaken here represent the first spectral-line polarisation observations with the Mt Pleasant 26m telescope and in Section 2 below we outline in detail the calibration process undertaken.  Rather than solving all the polarisation properties of the masers simultaneously it was more efficient to used different methods to determine the linear and circular polarisation properties independently.  The technique we applied to determine the linear polarisation (Stokes Q and U) assumes that there is no circularly polarized emission (Stokes V).  Although this is not strictly true, the level of circular polarisation observed in 6.7 GHz methanol masers is sufficiently small that it has no effect on our ability to measure linear polarisation at present (other measurement uncertainties play a much greater role).  

There have been several other published methods for calibrating a single dish radio telescope for polarisation observations. The approach used in this work for linear polarisation is similar to that used by \citet{Cenn09}. The main difference is that a highly linearly polarised calibrator source is not needed, calibration can be performed with just unpolarised sources. This is achieved by only expanding to first order and by applying some simplifying assumptions made based on the properties of 6.7 GHz methanol masers. The use of unpolarised sources as calibrators is a general result of \citet{Hamaker00}, but this paper details why and how this is applicable to these observations with the Mount Pleasant 26m. While some of the assumptions made in this work are known to be slightly inaccurate \citep{Straten04}, there are other sources of error that limit the accuracy of the results. Other methods include observations of sources over a wide range of parallactic angles such as the calibrations performed by \citet{Heiles01}, \citet{Johnston02} and \citet{Straten04}.

We have structured the paper in the following way.  Section 2 outlines the theoretical basis of the method we have used to measure the linear polarisation properties of the observed methanol masers.  Differences in the telescope, receiver systems and spectrometer systems at each telescope mean that there is no standard procedure for doing this and so we describe in detail the process used here.  Section 3 summarises the observations, and Section 4 describes the calibration observations and processing required to apply the theory outline in Section 2.  In Section 5 we discuss the results of our linear polarisation observations towards a sample of strong 6.7 GHz class II methanol masers.  In Section 6 we outline the procedures required to measure circular polarisation in spectral line sources and in Section 7 we describe the results of our circular polarisation observations.  Section 8 presents the conclusions.


\section{Spectral line linear polarisation measurement with the Mt Pleasant 26m}

In order to make linear polarisation measurements with any instrument it is necessary to relate the signal received by the processing hardware (in this case a digital autocorrelation spectrometer) with the signals emitted by the source of interest.  While the general process through which this is done is common to all systems the specific details of the system (e.g. telescope mount, receiver characteristics) mean that in practice it is an instrument specific procedure.  It is necessary to consider each section of the signal path from the emitting source through to the processing hardware and determine the effect of each section on the polarisation properties of the signal.  In this section we have undertaken that process for observations of 6.7 GHz methanol masers made with the University of Tasmania's 26m radio telescope at the Mt Pleasant observatory.  We have done this in the broader context of the mathematical framework of radio polarimetry for a single dish radio telescope, which (with appropriate modification) can be applied to observations with other receiver systems, or at other telescopes. The aim is to produce a 4x4 Mueller matrix in the Stokes frame that includes all the significant transformations that occur to the signal, including those due to the error in the telescope receiver feed. This will allow for calibration of the telescope for polarisation observations. The starting point for this is the mathematical framework set out by \citet{Hamaker96} which provides the connection between 2x2 Jones' matrices in the signal domain with 4x4 Mueller matrices in the coherency domain. This allows for the transformations that occur to the observed signal as it propagates through space and the telescope system to be described by a series of relatively simple 2x2 Jones matrices with their corresponding 4x4 Mueller matrices.  We consider each section of the signal path separately, starting from the emission of the radiation by the source through to the input signals to the processing hardware. Outlined in this section are these Mueller matrices for each section of the signal path and their combined product, the total Mueller matrix for the system.

\subsection{Faraday rotation}
As the emitted radiation from the source propagates towards the observer ionised material in the interstellar medium causes the linear polarisation to change position angle (Faraday rotation). The amount of rotation depends on the wavelength of the radiation and so for continuum sources this effect can be measured by observing the polarisation at a range of frequencies.  However, this method cannot in general be used for spectral line emission as it occurs at a single narrow frequency range.  In theory Faraday rotation can be measured for methanol masers, as they are known to produce cospatial emission from multiple transitions \citep{Menten92,Norris98}.  If it is assumed that the polarized component of this emission is the same then observations of both transitions can be used to determine the rotation measure along the line of sight from the observed maser.  Our original intention was to attempt this using observations of both the 6.7 and 12.2 GHz methanol transitions, however, technical difficulties with the 12.2 GHz receiver at the time of the observations meant that we were only able to make polarisation observations of the 6.7 GHz methanol line and hence the Faraday rotation could not be determined.  However, we do not expect this to significantly effect our results, as using models of the electron distribution in the Galaxy \citet{Dodson08} found that the Faraday rotatation is +0.8$^{\circ}$ at 6.7 GHz for the source G339.88-1.26. While this will vary from source to source the amount of Faraday rotation at this frequency is small enough to be negligible compared to other sources of uncertainty in our polarisation measurements.

\subsection{Coparallactic angle rotation}
As the source being observed moves across the sky it also rotates relative to the telescope receiver. Most radio telescopes have an azimuth-elevation mount, however the Mt Pleasant telescope is an X-Y mount telescope (with the X-axis aligned north-south).  For an X-Y mounted telescope this rotation is called coparallactic rotation (as opposed to parallatic rotation for an azimuth-elevation mount).  From spherical trigonometry the coparallactic angle rotation can be shown to be
\begin{equation}
\theta_p = 90^{\circ} + \arctan\left(\frac{\cos(h)}{\sin(h)\sin(\delta)}\right)
\end{equation}
Where $\theta_p$ is the coparallactic rotation angle, $h$ is the hour angle of the source and $\delta$ is the declination of the source. It should be noted that this is actually independent of the latitude of the observatory, because an X-Y telescope is aligned north-south and east-west.\\

Physical rotations, like the coparallactic angle, produce a two times angle rotation in the Stokes coordinates Q and U. Hence coparallactic rotation for a native circular receiver results in the following Mueller matrix
\begin{equation}
\mathbf{M}^{S}_{\mathrm{copara}}
=
\left(
\begin{array}{cccc}
1 & 0 & 0 & 0 \\
0 & \cos2\theta_p & \sin2\theta_p & 0 \\
0 & -\sin2\theta_p & cos2\theta_p & 0 \\
0 & &  0 & 1 \\
\end{array}
\right)
\end{equation}
Note that the $S$ superscript refers to the Stokes frame in the coherency domain.

\subsection{Receiver properties}
Firstly the receiver angle relative to the telescope axis has to be known. This is a physical property of the telescope and is constant unlike the time dependent coparallactic angle rotation. This rotation is described by the Mueller matrix $\mathbf{M}^{S}_{\mathrm{feed\;angle}}$ which has the same form as $\mathbf{M}^{S}_{\mathrm{copara}}$.\\

The type of receiver feed also has to be accounted for. For a native circular receiver this is typically done in one of two ways; either by a transposition of the Stokes vector from $\mbox{\boldmath$e$}^S = \{I,Q,U,V\}$ to $\mbox{\boldmath$e$}^S = \{I,V,Q,U\}$ or by a change in the transformation from the instrumental frame (correlator outputs) to the Stokes frame. In order to use the canonical Stokes vector $\mbox{\boldmath$e$}^S = \{I,Q,U,V\}$ the second approach will be adopted in this work.\\

\citet{Hamaker96b} give the transformation from the Stokes frame to the right-left circular coherency frame to be
\begin{equation}
\mathbf{S}
=\frac{1}{2}
\left(
\begin{array}{cccc}
1 & 0 & 0 & 1 \\
0 & 1 & \mathrm{i} & 0 \\
0 & 1 & -\mathrm{i} & 0 \\
1 & 0 & 0 & -1 \\
\end{array}
\right)
\end{equation}

\subsection{Imperfect receiver feed}
The error in the receiver has to be accounted for via the feed error Mueller matrix $\mathbf{M}^{S}_{\mathrm{error}}$. The framework described by \citet{Hamaker96} explains how to get from 2x2 Jones matrices representing transformations to the voltages to 4x4 Mueller matrices representing transformations to the Stokes parameters. This allows for a large reduction in the number of unknown parameters involved in the calibration process, depending on the Jones matrix used and the order of terms kept in the expansion.\\

The general Jones feed error matrix used by \citet{Heiles01} is
\begin{equation}
\mathbf{J}_{\mathrm{error}} = 
\left(
\begin{array}{cc}
1 & d_\mathrm{x}\\
\overline{d_\mathrm{y}} & 1\\
\end{array}
\right)
\end{equation}
Where $d_\mathrm{x}$ and $d_\mathrm{y}$ represent the complex leakage between the two dipoles in the receiver due to misalignment or non-circularity. $\overline{d_\mathrm{y}}$ is the complex conjugate of $d_\mathrm{y}$. The corresponding Mueller matrix $\mathbf{M}^{S}_{\mathrm{error}}$ is then given by
\begin{equation}
\mathbf{M}^{S}_{\mathrm{error}} = 
\mathbf{S}^{-1}(\mathbf{J}_{\mathrm{error}}\otimes\overline{\mathbf{J}_{\mathrm{error}}})\mathbf{S}
\end{equation}
Where $\otimes$ is the outer matrix product, defined in Appendix A of \citet{Hamaker96}.\\

Performing this expansion to first order in $d$ produces the following Mueller matrix in the Stokes frame
\begin{equation}
\mathbf{M}^{S}_{\mathrm{error}} = 
\left(
\begin{array}{cccc}
1 & x_1 + y_1 & x_2 + y_2 & 0\\
x_1 + y_1 & 1 &  0 & y_1 - x_1 \\
x_2 + y_2 & 0 & 1 & y_2 - x_2 \\
0 & x_1 - y_1 & x_2 - y_2 &  1 \\
\end{array}
\right)
\end{equation}
Where $d_x = x_1 + \mathrm{i} x_2$ and $d_y = y_1 + \mathrm{i} y_2$.\\

The elements of $\mathbf{M}^{S}_{\mathrm{error}}$ in the fourth column and fourth row represent leakages to and from Stokes V. These terms are due to the difference between $d_\mathrm{x}$ and $d_\mathrm{y}$ and represent the amount of ellipticity present in the circular receiver \citep{Heiles01}. These terms can be neglected in this case as Stokes V is expected to be insignificant due to the properties of the sources being observed. Without the out of phase couplings the error in the feed can be represented with just two parameters and the following Jones matrix
\begin{equation}
\mathbf{J}_{\mathrm{error}}  = 
\left(
\begin{array}{cc}
1 & d\\
\overline{d} & 1\\
\end{array}
\right)
\end{equation}
Where $d = d_1 + \mathrm{i} d_2$;  ${d_1 = \frac{1}{2}(x_1 + y_1)}$ and ${d_2 = \frac{1}{2}(x_2 + y_2)}$. This results in the following feed error matrix in the Stokes' domain
\begin{equation}
\mathbf{M}^{S}_{\mathrm{error}} = 
\left(
\begin{array}{cccc}
1 & 2 d_1 & 2 d_2 & 0 \\
2 d_1 & 1 &  0 & 0 \\
2 d_2 & 0 & 1 & 0 \\
0 & 0 & 0 &  1 \\
\end{array}
\right)
\end{equation}

\subsection{System properties}
The telescope has an intrinsic gain and system phase for each of the channels. This can be represented by the following Jones matrix 
\begin{equation}\label{Jonessystem}
\mathbf{J}_{\mathrm{sys}} = 
\left(
\begin{array}{cc}
g_1 & 0\\
0 & g_2 e^{\mathrm{i}\psi_{sys}}\\
\end{array}
\right)
\end{equation}
Where $g_1$ and $g_2$ are real and represent the voltage gains of the two polarisation channels. A noise diode connected to a short dipole, located inside the receiver, prior to the probes used to extract the incoming radiation is used to determine the system gains. The dipole attached to the noise diode produces linearly polarised radiation with roughly equal power received by both of the receiver probes. The strength of the noise diode is measured relative to sources of known flux density to determine the relative gains of each of the polarisation channels.\\

The phase difference $\psi_{sys}$ in Equation~\ref{Jonessystem} typically arises from differences in the electrical path length that the signals have to travel between the receiver and the correlator. This can be also dealt with by observing the correlated signal injected by the noise diode. In the presence of this strong, correlated signal, a difference in electrical path length produces a linear variation with frequency of the phase of the complex cross-correlation product between the two polarisations \citep{Heiles01}. A linear trend line can be fitted to this across the bandpass and the result subtracted from the phase of the on source cross-correlation observations. For a native circular receiver this subtraction is a rotation between Stokes Q and U. It should be noted that this process introduces a small error due to the probe with the noise diode attached radiating into the feed, and hence the feed error could potentially change the effective observed position angle of the probe. The emission from the noise diode probe is however highly linearly polarised, and so this error in position angle should be insignificant.\\

The use of a correlated noise signal allows for the Mueller matrix $\mathbf{M}^{S}_{\mathrm{sys}}$ arising from the Jones matrix $\mathbf{J}_{\mathrm{sys}}$ to be accounted for before further analysis of the data is attempted.

\subsection{Overall signal path}
The overall signal path and its transformations on the \textbf{in}coming Stokes vector to produce the \textbf{obs}erved Stokes vector can be represented as follows
\begin{equation}
\mbox{\boldmath$e$}^S_\mathrm{obs} = \mathbf{M}^{S}_{\mathrm{sys}} \mathbf{M}^{S}_{\mathrm{error}} \mathbf{M}^{S}_{\mathrm{feed\;angle}} \mathbf{M}^{S}_{\mathrm{copara}} \mbox{\boldmath$e$}^S_\mathrm{in}
\end{equation}
The constant rotation due to the feed angle can be absorbed into the rotation for the coparallactic angle. The relative gains and and system phase can be accounted for to deal with $\mathbf{M}^{S}_{\mathrm{sys}}$. The type of feed has already been accounted for by specifying $\mathbf{S}$.
\begin{equation}
\mbox{\boldmath$e$}^S_\mathrm{obs} = \mathbf{M}^{S}_{\mathrm{error}} \mathbf{M}^{S}_{\mathrm{copara}} \mbox{\boldmath$e$}^S_\mathrm{in}  
\end{equation}
Here $\mathbf{M}^{S}_{\mathrm{copara}}$ is the time dependent part of the total Mueller matrix of the system and $ \mathbf{M}^{S}_{\mathrm{error}}$ is the unknown time independent part representing the feed error that needs to be determined via calibration. These can be combined into a single Mueller matrix $\mathbf{M}^{S}$ describing the system as a whole.

\begin{equation} \label{MS}
\mbox{\boldmath$e$}^S_\mathrm{obs} = \mathbf{M}^{S}\mbox{\boldmath$e$}^S_\mathrm{in}
\end{equation}
\begin{equation}\label{mueller}
\mathbf{M}^{S}
=
\left(
\begin{array}{cccc}
1 & K_1 & K_2 & 0\\
2d_1  & \cos(2\theta_p) & \sin(2\theta_p) & 0 \\
2d_2 & -\sin(2\theta_p) & \cos(2\theta_p) & 0 \\
0 & 0 & 0 & 1\\
\end{array}
\right)
\end{equation}
Where ${K_1= 2d_1\cos(2\theta_p)-2d_2\sin(2\theta_p)}$ and \\
${K_2 = 2d_1\sin(2\theta_p)+2d_2\cos(2\theta_p)}$.\\

$\mathbf{M}^{S}$ represents the total transformations that occur to the incoming signal after taking into account the type of telescope feed and calibrating for the system phase. This description of the telescope system can now be used to determine what the unknown feed error is and calibrate the telescope for linear polarisation observations.

\section{Observations}
Observations were conducted using the University of Tasmania's 26m radio telescope located at Mount Pleasant. This telescope has an X-Y mount, native circular receivers at 6.7 GHz, a main beam FWHM of 7.5 arcminutes at 6.7 GHz and a typical pointing accuracy of 30 arcseconds. All of the observed sources have accurate positions given by \citet{Caswell09} and \citet{Caswell10}. These sources are compact and have negligible structure at single dish resolution, meaning that off-axis polarisation properties are insignificant. \\

A 2-bit digital auto-correlation spectrometer was used to correlate the data. For linear polarisation observations a 4 MHz bandwidth with 2048 spectral channels was used. This  produced 4 outputs; the left and right hand circular auto-correlations as well as the real and imaginary parts of the left-right cross correlation. For circular polarisation observations a 4 MHz bandwidth with 4096 spectral channels was used, this produced the left and right hand circular auto-correlations only.\\

The majority of the observations reported in this paper were carried out over the period 2010 September 28-30. The observing strategy involved repeated observations of the strong 6.7 GHz methanol maser sources G323.74-0.26, G339.88-1.26 and G351.42+0.64 over a range of hour angles.  Observations of the continuum source Virgo A for calibration purposes and of several other southern 6.7 GHz methanol masers were also undertaken. On 2011 February 12 further observations were undertaken, repeating the calibration observations of Virgo A and a few of the 6.7 GHz methanol maser sources observed in September 2010. In addition we also made repeated calibration observations of the strong continuum source PKS B\,1921-293 (another largely unpolarised source) to perform additional tests on the repeatability, accuracy and reliability of the calibration procedures.\\

Observations of the masers were made by first making a short off-source reference pointing with the noise diode on to allow correction for $\psi_{sys}$, followed by another longer off-source reference observation with the noise diode off, and then an on-source observation (with the noise diode off) of the same duration. The shape of the cumulative receiver and correlator bandpasses were corrected by subtracting the off-source reference observations from the on-source, rather than the more typical division method, because the cross-correlation product was not always greater than zero.\\

A follow up circular polarisation observation of G9.62+0.20 was made during August 2011, this consisted of a total of 4 hours on-source, 4 hours off-source; split into 10 minute alternating scans. 

\section{Determining the feed error}
There is no single preferred method for measuring the feed error and hence obtaining the total Mueller matrix (Equation~\ref{mueller}) for the observing system. One method which has been used \citep{Heiles01,Johnston02} involves observing a partially polarised source over a wide range of hour angles. This produces rotation in the observed source relative to the telescope feed due to parallactic (alt-az mount) or coparallactic (X-Y mount) rotation. Sine functions are then fitted to the fractional Stokes parameters of the source against the parallatic (or for an X-Y telescope, the coparallactic) angle. Fractional Stokes parameters are used due to there typically being variations in the total system gain with elevation. This method is especially useful when using native linear receivers and/or attempting to use more terms in the feed error matrix. The problem with this method of calibration is that it requires observations over a large range of hour angles, requiring significant amounts of telescope time.\\

Another method which can be used to determine the feed error matrix is to observe one or more unpolarised sources. This requires much less observing time, but since Stokes V is dominated by errors in relative gain for a native circular receiver only two products can be used, fractional Stokes Q and U. This produces 2 equations and so cannot be applied to systems with more than 2 unknown parameters in the feed error matrix. It does however allow for independent calibration of each spectral channel across the entire bandpass (in contrast to the previous method when applied to maser emission). \\

In this section we outline how we used these two techniques to determine the total Mueller matrix (Equation~\ref{mueller}) for our system.  This is then used to determine the feed error of the 6.7 GHz receiver on the Mt Pleasant 26m telescope. Observations of unpolarised sources were used as the primary calibration method in this work, however observations over a range of coparallactic angles were also completed to compare the two methods. \\

\subsection{Partially polarised source}
When calibrating using a partially polarised maser source, the incoming circular polarisation and hence Stokes V is assumed to be zero, resulting in the following Stokes vector
\begin{equation}
\mbox{\boldmath$e$}^S_\mathrm{in}=
\left(
\begin{array}{c}
I_\mathrm{in}\\
Q_\mathrm{in}\\
U_\mathrm{in}\\
0\\
\end{array}
\right)
\end{equation}
After applying the total Mueller matrix for the system $\mathbf{M}^{S}$ in Equation \ref{mueller} to this vector the observed fractional Stokes parameters can be expressed in terms of coparallactic angle, the feed error and the source Stokes parameters as follows
\begin{align}
q_\mathrm{obs} &= \frac{Q_\mathrm{obs}}{I_\mathrm{obs}}=\frac{2d_1 I_\mathrm{in} + \cos(2\theta_p) Q_\mathrm{in} + \sin(2\theta_p) U_\mathrm{in}}{I_\mathrm{in}+K_1 Q_\mathrm{in}+K_2 U_\mathrm{in}}\\
u_\mathrm{obs} &= \frac{U_\mathrm{obs}}{I_\mathrm{obs}}=\frac{2d_2 I_\mathrm{in} - \sin(2\theta_p) Q_\mathrm{in} + cos(2\theta_p) U_\mathrm{in}}{I_\mathrm{in}+K_1 Q_\mathrm{in}+K_2 U_\mathrm{in}}
\end{align}
As the feed error matrix was only expanded to first order, the approximation can be made that $I_\mathrm{obs} \approx I_\mathrm{in}$ and hence the expressions reduce to
\begin{equation} \label{eqcopara}
q_\mathrm{obs} = 2d_1 + \sqrt{q_\mathrm{in}^2+ u_\mathrm{in}^2}\cos\left(2\theta_p+\Phi_1\right)
\end{equation}
\begin{equation}
u_\mathrm{obs} = 2d_2 + \sqrt{q_\mathrm{in}^2+ u_\mathrm{in}^2}\cos\left(2\theta_p+\Phi_2\right)
\end{equation}
Where $\Phi_1$ and $\Phi_2$ are phase shifts depending on $Q_{in}$ and $U_{in}$.\\

\begin{figure}[h!]
\begin{center}
\includegraphics[scale = 0.35]{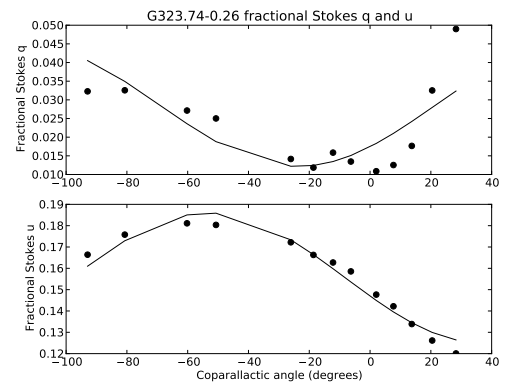}
\caption{Fractional Stokes q and u against coparallactic angle with fit for G323.74-0.26}\label{coparafit1}
\end{center}
\end{figure}

\begin{figure}[h]
\begin{center}
\includegraphics[scale = 0.35]{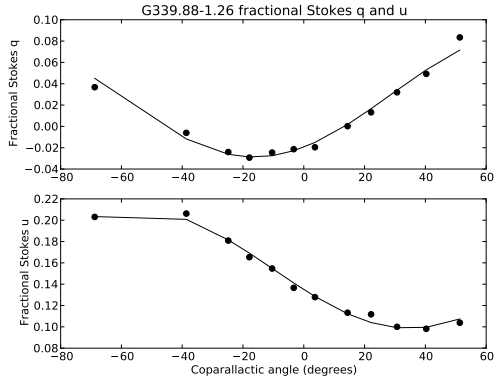}
\caption{Fractional Stokes q and u against coparallactic angle with fit for G339.88-1.26}\label{coparafit2}
\end{center}
\end{figure}

\begin{figure}[h]
\begin{center}
\includegraphics[scale = 0.35]{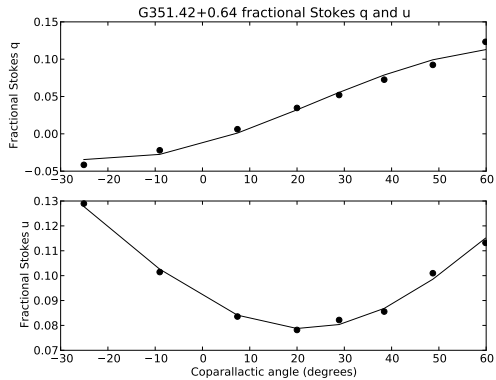}
\caption{Fractional Stokes q and u against coparallactic angle with fit for G351.42+0.64}\label{coparafit3}
\end{center}
\end{figure}

Observations of three strong (peak flux density $>$ 1000 Jy) 6.7 GHz methanol masers G323.74-0.26, G339.88-1.26 and G351.42+0.64 were completed over a range of coparallactic angles over the course of a day. For each source we averaged the emission over a velocity range covering the peak exhibiting the greatest linear polarisation to produce a fractional Stokes q and u value. Non-linear least squares fitting was then used to fit a function of the form ${Y = A\cos(2\theta_p+B)+C}$ to the fractional Stokes q and u plots against coparallactic angle. While the source Stokes parameters are unknown at this stage, the A coefficient from this fit gives an estimate of the fractional linear polarisation present. The C coefficients from the fits are equal to the $2d_1$ and $2d_2$ terms from the feed error matrix. The fits to the observations are shown in Figures \ref{coparafit1}--\ref{coparafit3} and the derived error matrix terms are given in Table \ref{coparafit} along with their formal uncertainties from the fits. The errors due to radiometer noise for each point in Figures \ref{coparafit1}--\ref{coparafit3} are smaller than the size of the points used and hence are not shown.

It should be noted that while the fit to fractional Stokes q and u for G323.74-0.26 in Figure \ref{coparafit1} does not appear as to be as good as the fits seen in Figures \ref{coparafit2} and \ref{coparafit3} the errors in the fits are actually very similar. The error just appears large in Figure \ref{coparafit1} due to lower fractional linear polarisation in G323.74-0.26 (around $\sim2\%$ compared to an average of ${\sim6\%}$ for the other two sources). Another point to note is that while the range of coparallactic angles observed is less than $180^\circ$, the fits are restricted to a known angular frequency and so the length of the data set is sufficient to obtain accurate fits. 

\begin{table}[h]
\begin{center}
\caption{Coparallactic angle fit results}\label{coparafit}
2010 September 28-30\\
\begin{tabular}{ccc}
\hline 
Source & $2d_1$ & $2d_2$ \\
\hline
G323.74$-$0.26 & 0.028$\pm$0.017 & 0.156$\pm$0.006 \\
G339.88$-$1.26 & 0.030$\pm$0.012 & 0.154$\pm$0.004 \\
G351.42+0.64 & 0.040$\pm$0.010 & 0.126$\pm$0.002 \\
\hline
Mean & $0.033\pm0.007$ & $0.145\pm0.017$\\
\end{tabular}
\end{center}
\end{table}

\subsection{Unpolarised source calibration}
When calibrating polarisation using an unpolarised source the incoming Stokes Q, U and V are assumed to all be zero.
\begin{equation}
\mbox{\boldmath$e$}^S_\mathrm{in}=
\left(
\begin{array}{c}
I_\mathrm{in}\\
0\\
0\\
0\\
\end{array}
\right)
\end{equation}
The resulting observed fractional Stokes q and u are then simply given by
\begin{align}
q_\mathrm{obs} &= \frac{Q_\mathrm{obs}}{I_\mathrm{obs}}=2d_1\\
u_\mathrm{obs} &= \frac{U_\mathrm{obs}}{I_\mathrm{obs}}=2d_2
\end{align}
A continuum calibrator source, such as Virgo A, can then be observed to allow for calibration across the entire bandpass, from a single on-off telescope pointing. A linear trend line is fitted to the fractional Stokes q and u across the bandpass to reduce the effect of noise on the data, this trend line is then used to calibrate each channel individually. The means can work just as well however as the slope across the bandpass was typically very small. Observations of both Virgo A and PKS B\,1921$-$293 were made, with repeated observations of PKS B\,1921$-$293 undertaken to test the repeatability of the calibration method.  Extragalactic radio sources such as Virgo A and PKS B\,1921$-$293 do sometimes exhibit linear polarisation at the level of a few percent in their core components.  In single dish observations, the spatial blending within a single beam of these compact polarised components, is further diluted when combined with the largely unpolarised kpc scale emission and together this reduces the fractional linear polarisation significantly.  To first order these sources can be considered unpolarised to our system.  

\begin{table}[h]
\begin{center}
\caption{Unpolarised source calibration results}\label{unpol}
2010 September 28-30\\
\begin{tabular}{ccc}
\hline 
Source & $2d_1$ & $2d_2$ \\
\hline
Virgo & 0.0442$\pm$0.0005 & 0.1383$\pm$0.0004\\
\hline
\end{tabular}\\[\baselineskip]
2011 February 12\\
\begin{tabular}{ccc}
\hline 
Source & $2d_1$ & $2d_2$ \\
\hline
Virgo A & 0.0795$\pm$0.0004 & 0.0676$\pm$0.0004\\
PKS B\,1921$-$293a & 0.093$\pm$0.001 & 0.056$\pm$0.001\\
PKS B\,1921$-$293b & 0.099$\pm$0.001 & 0.065$\pm$0.001\\
\textbf{PKS B\,1921$-$293c} & \textbf{-0.177$\pm$0.001} & \textbf{0.192$\pm$0.001}\\
\hline
Mean & $0.090\pm0.010$ & $0.063\pm0.006$\\
\end{tabular}
\end{center}
\end{table}

The errors quoted in Table \ref{unpol} are the standard errors due to radiometer noise. The uncertainty in the mean is taken as the standard deviation of the values of 2$d_1$ and 2$d_2$ determined. The final result has an uncertainty significantly larger than that of the individual observations, indicating that there is another unknown source of error besides radiometer noise affecting the results. The source of this error is possibly physical in nature, based on the direction the telescope is pointing. The Mount Pleasant 26m is an X-Y mount telescope, unlike alt-az mounts this results in differing orientations of the receiver platform relative to the ground and hence differing gravitational stresses on the receiver. These stresses could cause slight rotations or deformations of the receiver setup and slightly change the polarisation characteristics of the telescope.\\

In Table \ref{unpol} the a, b and c for the source PKS B\,1921$-$293 refer to observations made at the beginning, middle and end of the observation run. The observation at the end (labeled PKS B\,1921$-$293c in the table), produced very different values for $d_1$ and $d_2$.  It is not clear why this occurred and so this result has been included in the table, but was not included in the mean or further analysis. If it is due to the previously mentioned physical effects then it would indicate they can be far larger than expected. Its presence does however indicate that multiple calibration observations should be made so as to avoid possible individual anomalous results. Another point of interest is the fact that the measured $d_1$ and $d_2$ changed significantly in between September 2010 and February 2011.  Although the basic observing system and strategy were the same for the two sessions individual components in the system are likely to have been changed.  Further work should be undertaken to monitor the magnitude and rate of change in the polarisation calibration of the system.\\

As can be seen comparing the results for 2010 September in Tables \ref{coparafit} and \ref{unpol}, there is fairly good agreement between the values determined for $d_1$ and $d_2$ between the two methods. This indicates that the observation of unpolarised continuum sources is a valid calibration method for this type of telescope, though multiple calibration observations should be made to ensure accurate and reliable results.

\section{Linear polarisation results and discussion} \label{linear}
\begin{figure}[h!]
\begin{center}
\includegraphics[scale = 0.35]{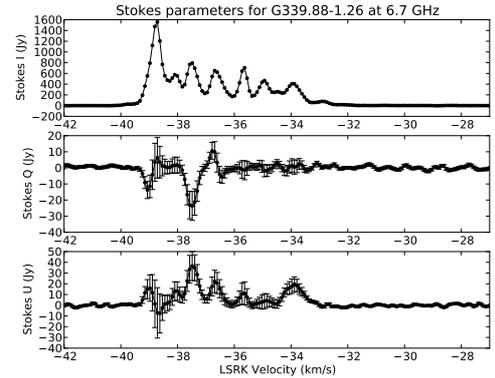}
\caption{Stokes I, Q and U for G339.88-1.26, observations made during September 2010}\label{g339a}
\end{center}
\end{figure}

\begin{figure}[h!]
\begin{center}
\includegraphics[scale = 0.35]{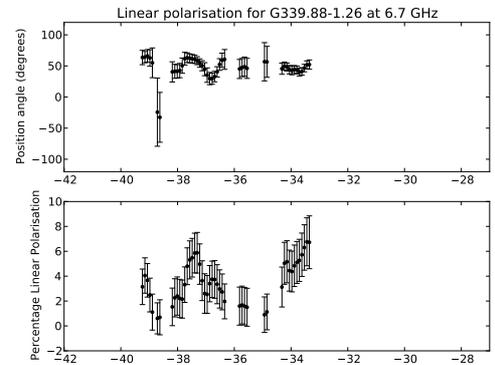}
\caption{Percentage linear polarisation and position angle for G339.88-1.26, observations made during September 2010}\label{g339b}
\end{center}
\end{figure}

The linear polarisation observations conducted are shown in the appendix. These are divided up into three groups; the sources which were observed in both September 2010 and February 2011, the sources with significant linear polarisation observed only in September 2010 and the sources observed in September 2010 which do not show significant linear polarisation. Significant linear polarisation in this case is defined as greater than 5 times the RMS noise level in the Stokes q and u observations, allowing for a meaningful plot of position angle and fractional linear polarisation. Tables \ref{septlin} and \ref{feblin} summarise these results for sources with significant linear polarisation. Source coordinates in these tables are taken from the methanol maser catalogue in \citet{Caswell09}.  Example spectra for G339.88-1.26 are shown in Figures \ref{g339a} and \ref{g339b}.

\begin{table*}[h!]
\begin{center}
\caption{Linear polarisation results, September 2010}\label{septlin}
\begin{tabular}{ccccccccc}
Source & RA(2000) & Dec(2000) & V$_{pk}$ (km s$^{-1}$) & S$_{pk}$ (Jy) & \% Lin Pol$_{pk}$ & Angle$^\circ_{pk}$ & Angle range$^\circ$ \\
\hline
G309.92+0.48 & 13 50 41.8  & -61 35 10 & -59.7 & 980 & 5.9$\pm$1.3 & 10$\pm$8 & -15--29\\
G316.64-0.09 & 14 44 18.5 & -59 55 12 & -20.4 & 120 & 4.7$\pm$1.8 & 51$\pm$8 & 45--52\\
G318.95-0.20 & 15 00 55.3 & -58 58 53 & -34.7 & 560 & 2.8$\pm$1.2 & 20$\pm$14 & -33--22\\
G322.16+0.64 & 15 18 34.6 & -56 38 25 & -63.0 & 256 & 2.5$\pm$1.6 & 45$\pm$11 & 43--60\\
G323.74-0.26 & 15 31 45.5 & -56 30 50 & -50.5 & 2500 & 2.3$\pm$1.5 & 49$\pm$11 & -27--76 \\
G328.81+0.63 & 15 55 48.7  & -52 43 06 & -44.5 & 280 & 5.1$\pm$1.6 & -30$\pm$7 & -31--18\\
G329.03-0.21 & 16 00 31.8 & -53 12 50 & -37.0 & 103 & 4.1$\pm$1.8 & 39$\pm$9 & 34--39\\
G335.79+0.17 & 16 29 47.3 & -48 15 52 & -47.7 & 214 & 5.6$\pm$1.3 & 12$\pm$8 & 5--33\\
G339.88-1.26 & 16 52 04.7 & -46 08 34 & -38.7 & 1560 & 0.6$\pm$1.2 & -23$\pm$56 & -32--67\\
G345.01+1.79 & 16 56 47.6 & -40 14 26 & -22.2 & 306 & 4.6$\pm$1.2 & 14$\pm$10 & 5--25\\
G351.42+0.64 & 17 20 53.4 & -35 47 01 & -10.4 & 3430 & 9.0$\pm$2.0 & 48$\pm$5 & -89--73\\
G353.41-0.36 & 17 30 26.2 & -34 41 46 & -20.4 & 84 & 7.7$\pm$1.7 & -5$\pm$8 & -5--1\\
G9.62+0.20 & 18 06 14.7 & -20 31 32 & 1.3 & 5340 & 2.3$\pm$1.2 & -23$\pm$15 & -23--11\\
\end{tabular}
\end{center}
\end{table*}

\begin{table*}[h!]
\begin{center}
\caption{Linear polarisation results, February 2011}\label{feblin}
\begin{tabular}{ccccccccc}
Source & RA(2000) & Dec(2000) & V$_{pk}$ (km s$^{-1}$) & S$_{pk}$ (Jy) & \% Lin Pol$_{pk}$ & Angle$^\circ_{pk}$ & Angle range$^\circ$ \\
\hline
G309.92+0.48 & 13 50 41.8  & -61 35 10 & -59.7 & 1090 & 4.9$\pm$1.0 & 19$\pm$6 & -16--44\\
G323.74-0.26 & 15 31 45.5 & -56 30 50 & -50.5 & 3870 & 3.4$\pm$0.8 & 30$\pm$8 & -13--35\\
G339.88-1.26 & 16 52 04.7 & -46 08 34 & -38.7 & 1660 & 1.1$\pm$1.0 & 0$\pm$15 & -3--55\\
G345.01+1.79 & 16 56 47.6 & -40 14 26 & -22.2 & 336 & 3.8$\pm$1.0 & 18$\pm$7 & 9--30\\
G351.42+0.64 & 17 20 53.4 & -35 47 01 & -10.4 & 3810 & 9.4$\pm$1.2 & 56$\pm$4 & -65--78\\
G9.62+0.20 & 18 06 14.7 & -20 31 32 & 1.3 & 6500 & 2.3$\pm$0.8 & -28$\pm$12 & -30--10\\
\end{tabular}
\end{center}
\end{table*}

\subsection{Comparison with other results}
Comparisons of several of the spectra were made with observations made using the ATCA in 1999 (Simon Ellingsen, Private Communication). Many of the features agreed to within one or two sigma for G9.62+0.20, G339.88$-$1.26, G345.01+1.79 and G351.42+0.64, which is a reasonable result considering the significant changes which have occurred in the Stokes I profile in the time interval between the two sets of observations. \citet{Dodson08} also made polarisation observations of G339.88$-$1.26 at 6.7 GHz and these again are in approximate agreement with the observations made in this work.\\

Observations were undertaken in February 2011 to compare with the observations made in September 2010. As can be seen from the appendix, these polarisation spectra of the same source from the two sessions agree closely. This gives us confidence that the calibration method is working correctly, even if the feed error properties of the telescope changed in between the two sets of observations. 

\subsection{Error estimation}
Three sources of error were included in our estimate of the total error in the Stokes Q and U values determined. The first source of error considered is that due to the first order expansion in the feed error matrix.  The first order expansion assumes that there is a small total leakage from Stokes I to Stokes Q and U. This was characterised by 
\begin{equation}
E_1 = \sqrt{Q^2+U^2}\sqrt{(2d_1)^2+(2d_2)^2}
\end{equation}
The second source of error is that in the feed error matrix term.  To estimate the uncertainty in each of these terms we used the distribution of values from the multiple independent measurements of these terms.
\begin{equation}
E_{2Q} = I  \Delta(2d_1)
\end{equation}
\begin{equation}
E_{2U} = I  \Delta(2d_2)
\end{equation}
The final source of error we have considered is the measured noise level in the Stokes Q and U spectra. This had to be included so that parts of the spectrum with no Stokes I signal did not confuse the final polarisation spectra.
\begin{equation}
E_3 = \textrm{RMS variation in Stokes Q and U}
\end{equation}
These errors were combined in quadrature to obtain the total estimated error in Stokes Q and U.
\begin{equation}
\Delta Q = \sqrt{E_1^2 + E_{2Q}^2 + E_3^2}
\end{equation} 
\begin{equation}
\Delta U = \sqrt{E_1^2 + E_{2U}^2 + E_3^2}
\end{equation}
Uncertainty in the fractional linear polarisation and position angle were then determined by standard error propagation from the estimated Stokes Q and U errors.

\subsection{Discussion}
Recent work by \citet{Ellingsen07,Breen10,Breen11} and others, has attempted to find an evolutionary timeline for the different maser species and transitions.  In particular, \citet{Breen10} suggests that the more luminous 6.7 GHz methanol masers are associated with less dense dust clumps than the lower luminosity masers.  The magnetic field is expected to scale proportional to the gas density and so we might expect to see some relationship between the maser luminosity and the polarisation properties (since these are thought to be produced by the magnetic field).  Using the kinematic distance to determine the luminosity of the 6.7 GHz masers we observed we looked to see if there is a relationship between the fractional linear polarisation and the luminosity of the maser source.  No such relationship was found for our sample of 6.7 GHz methanol masers.  However, the primary selection criteria for the masers we targeted was that they have a peak flux density in excess of several hundred Jy (i.e. they are very bright masers). Further study involving weaker, intrinsically less luminous masers should be undertaken to definitively test whether there is any relationship between the polarisation properties and the maser luminosity.\\

The combination of comparisons with polarisation observations of 6.7 GHz methanol masers made with other telescopes and the checks for repeatability and self-consistency between different approaches of the Mt Pleasant observations, demonstrates that we are able to use this system to reliably measure the linear polarisation of masers.  The faster and easier calibration method using unpolarised sources developed in this work appears consistent with the more commonly used approach of observing a partially polarised source over a wide range of hour angles. This development will hopefully allow further linear polarisation observations of maser emission to be conducted with this telescope, or any other single dish radio telescope with native circular receivers.

\section{Circular polarisation}
Observations of circular polarisation with native circular receivers are highly susceptible to errors in relative gain between the receivers. To deal with this the system gains have to be handled carefully. This is usually done by first assuming that any observed Stokes V is due to Zeeman splitting, typically due to a line-of-sight magnetic field. After making this assumption there are two main methods for determining the magnitude of this splitting covered in the literature, S-curve fitting and the running cross-correlation.

\subsection{S-curve fitting}
S-curve fitting involves fitting Stokes I and the derivative of Stokes I ($\mathrm{dI}/\mathrm{d}v$) to the raw Stokes V spectra, which is the right hand circular autocorrelation minus the left hand circular autocorrelation. The form of the expression fitted is as follows
\begin{equation}
V=a\mathrm{I}+b\frac{\mathrm{dI}}{\mathrm{d}v}
\end{equation}
Where $a$ and $b$ are the constants determined via the fitting. If the circular polarisation is due to Zeeman splitting then the part of Stokes V proportional to Stokes I is expected to be due to errors in the relative gain, while the part of Stokes V proportional to the derivative of Stokes I is due to the Zeeman splitting. \cite{Vlemmings02} looked at synthetic Stokes V spectra and derived the following expression for magnetic field strength based on applying S-curve fitting to a Stokes V spectra.
\begin{equation}
\frac{V_\mathrm{max}-V_\mathrm{min}}{I_\mathrm{max}} = \frac{2 A_{F-F'} B_{||}}{\Delta v_\mathrm{L}}
\end{equation} 
Where $V_\mathrm{max}$ and $V_\mathrm{min}$ are the maximum and minimum of the fitted Stokes V, $\Delta v_\mathrm{L}$ is the full width half-maximum (FWHM) of Stokes I and $I_\mathrm{max}$ is the peak flux density of the maser feature. $B_{||}$ is the line of sight magnetic field strength and $A_{F-F'}$ is the Zeeman splitting coefficient for the line observed.

This method produces a Stokes V spectrum as well as a fitted derivative of the Stokes I to the Stokes V, but requires the FWHM of the feature in Stokes I to determine the Zeeman splitting. This can prove problematic in complex sources where the spectral peaks are not Gaussian and/or overlap in velocity; causing difficulty fitting them and blending resulting in reduction in the magnitude of the Stokes V detected. 

\subsection{Running cross-correlation}
Another method to determine the amount of Zeeman splitting directly without first forming a Stokes V profile is the running cross-correlation. The basics for performing a cross-correlation to determine the amount of Zeeman splitting are introduced by \citet{Modjaz05}. The process involves forming three products between the left hand circular and right hand circular autocorrelations at $+$1, 0 and $-$1 velocity channels of lag. As methanol is diamagnetic it exhibits weak Zeeman splitting and so the amount of splitting will typically be much less than the channel width. A quadratic can be fitted to these lags to determine how far in velocity the left and right hand autocorrelations are shifted relative to each other. The Zeeman splitting coefficient of the line can then be used to determine the line of sight magnetic field strength. If $x$ is the velocity shift between left and right hand circular polarisations the line of sight magnetic field is then given by \citet{Modjaz05} to be

\begin{equation} \label{ccsplit}
B_{||} = \frac{x}{\sqrt{2}A_{F-F'}}
\end{equation}

\citet{Vlemmings08} extended the method of \citet{Modjaz05} to use small velocity windows that are cross-correlated, these windows are moved across the spectra producing a running cross-correlation. This allows for variation in the magnetic field strength across the source to be determined, although smoothed depending on the size of the windows used. \citet{Vlemmings08} used 3 kms$^{-1}$ wide windows, as noise considerations limited the use of smaller windows. In this work however the only feature that provided a high enough signal to noise ratio to perform this analysis on was the main peak of G9.62+0.20 after an 8 hour integration. A 1.6 kms$^{-1}$ wide window centered at the main peak was used, this avoided the emission from the secondary feature.

\subsection{Zeeman splitting coefficient}
The Zeeman splitting parameter for the 6.7 GHz methanol transition has not been determined experimentally. Laboratory work has been done by \citet{Jen51} to determine the Lande-$g$ factor of the 25 GHz methanol lines, this can be used to estimate the splitting coefficient for the 6.7 GHz line. \citet{Vlemmings08} gives an estimate of the splitting coefficient of the 6.7 GHz methanol line to be 0.049 km s$^{-1} \mathrm{G}^{-1}$. It has recently emerged that an error was made in these calculations \citep{Fish11,Vlemmings11} and the correct result is an order of magnitude smaller. As a result of this we have repeated the calculations of the Zeeman splitting factor and obtain a value of 0.0048 km s$^{-1} \mathrm{G}^{-1}$ (consistent with \citeauthor{Fish11,Vlemmings11}), this is the value used in this work.  Section 3 of \citet{Vlemmings11} gives a detailed discussion of the current state of knowledge about this important parameter.  At present it would seem prudent to take the magnetic field estimates obtained in this and other work based on methanol maser observations as order of magnitude estimates, which may change if a better Zeeman splitting coefficient is determined for the 6.7 GHz methanol line.

\subsection{Beam squint}
Offsets in the pointing between the left and right hand circular receivers can produce false Stokes V for extended sources. Most emission from typical maser sources is however contained within a very small region, less than 100 mas \citep{Caswell97,Minier00}. This limits the false Stokes V due to beam squint to be typically much less than the true Stokes V, except when multiple sources are within the same beam. Comparisons with high resolution interferometer maps of the sources can help determine whether the Stokes V profile is due to beam squint. For example observations of G351.42+0.64 (NGC6334F) produced a profile consistent with significant beam squint due to strong emission in this source being present  over an angular extent of several arcseconds.  This is particularly an issue in this source as the emission from the two regions overlaps in velocity and is of comparable strength \citep{Ellingsen96}.  G9.62+0.20 is also known to have two sites of emission separated by around 13 arcseconds \citep{Caswell09}, however, in this case there is no overlap in velocity between the two sites and the emission at the offset site is much weaker.

\subsection{Other sources of error}
It should be noted that the approaches used to determine circular polarisation in this paper assume a perfect receiver. Non-circularity in the receiver feed will produce leakages from other Stokes products into Stokes V. Leakages from Stokes I are equivalent to errors in the relative gain and hence not an issue. Leakages from Stokes Q and U could potentially produce some false circular polarisation but this would typically not be in a form mimicking the derivative of the Stokes I spectrum which might be mistaken for Zeeman splitting. The results of Section \ref{linear} also show that the magnitude of Stokes Q and U detected in 6.7 GHz methanol masers would only produce measurable Stokes V if there were very high leakage.

\section{Circular polarisation results and discussion}
\begin{figure}[h]
\begin{center}
\includegraphics[scale = 0.41]{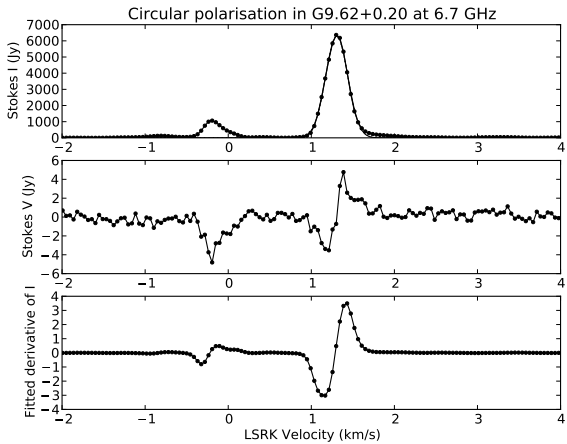}
\caption{S-curve fitting for G9.62+0.20}\label{scurve}
\end{center}
\end{figure}

Both S-curve fitting and the cross-correlation method were applied to spectra from several sources. Most sources however did not produce a high enough signal to noise ratio in the observation time to give meaningful Zeeman splitting estimates. G9.62+0.20 was the only exception to this, with a line of sight magnetic field strength of 35$\pm$7 mG via both the S-curve fitting and cross-correlation methods. The Stokes V spectra produced via S-curve fitting can be seen in Figure \ref{scurve}. Both of these methods were applied around the main peak only, avoiding the secondary emission. The Stokes V spectra for the secondary peak in Figure \ref{scurve} is typical of beam-squint, which is expected as the emission region is slightly off-set from the main region by about 13 arcseconds \citep{Caswell09}.

\subsection{Synthetic spectra simulations}
Monte Carlo simulations of synthetic spectra for G9.62+0.20 were performed to estimate the error in the magnetic field strengths determined. These simulations involved first generating two Gaussians to represent the left and right hand circular autocorrelations.  These had the same width and half the maximum of the Stokes I profile for G9.62+0.20 and were split in velocity by the shift determined via applying the cross-correlation method to the observed spectra. Simulated radiometer noise was then added to these, proportional to the system temperature in each channel based on the observations. The sum of these was then used as the simulated Stokes I profile. The noise added was set so that the Stokes I baseline produced had the same noise level as observed in the Stokes I profile. The channel spacing in velocity was set to that of the observations, $\sim$0.044 kms$^{-1}$. The velocities of the channels relative to the generated spectra were also shifted randomly to reduce any effect under sampling may have had on the results. These synthetic spectra were then put through the S-curve fitting and cross-correlation methods to estimate the magnetic field strength. This was repeated 100 times and the standard deviation of the results was used as the error in the magnetic field estimates.

\subsection{Comparison with other studies}
\citet{Vlemmings09} looked at the Zeeman splitting in G9.62+0.20 and found an average line of sight magnetic field strength of $11.0\pm2.2$ mG for the main feature, with a similar result for the secondary feature. This was done however with the Zeeman splitting coefficient of 0.049 km s$^{-1} \mathrm{G}^{-1}$. If the Zeeman splitting coefficient used in this work was applied then the field strength determined would be 112$\pm$22 mG, much larger than the value obtained in this work. The spectral resolution used in this work was slightly higher than what was used by \citet{Vlemmings09}, by a factor of 1.25. This should have resulted in better sampling of the very narrow features of the S-curve spectra, resulting in an effective increase in detected magnetic field strength, rather than a decrease. The Zeeman splitting in G9.62+0.20 is known to be variable however, as discussed by \citet{Vlemmings09}, so this may be the cause of the discrepancy between the determined magnetic field strengths.

\subsection{Discussion}
While the circular polarisation results in this work are limited, they do at least show it is possible to make circular polarisation observations of methanol masers with small single dish telescopes. Further studies of weaker masers using a 26m dish would involve many hours or days of observation for a single source. In practice large aperture single dish telescopes are required to make circular polarisation of all but the strongest methanol masers.

\section*{Acknowledgments} 
We would like to thank Wouter Vlemmings for informative discussions on circular polarisation observations of methanol masers and the Zeeman splitting factor.

\clearpage

\section*{Polarisation Spectra}

\begin{minipage}{\textwidth}
\subsection*{Linear polarisation comparison spectra}
These sources were observed on both 2011 February 12 and 2010 September 28-30, to test the repeatability of the calibration method. The September observations for G339.88-1.26 appear in Figures \ref{g339a} and \ref{g339b}.
\end{minipage}

\begin{tabular}{cc}
\begin{sideways}
2011 February 12
\end{sideways}
\includegraphics[scale = 0.35]{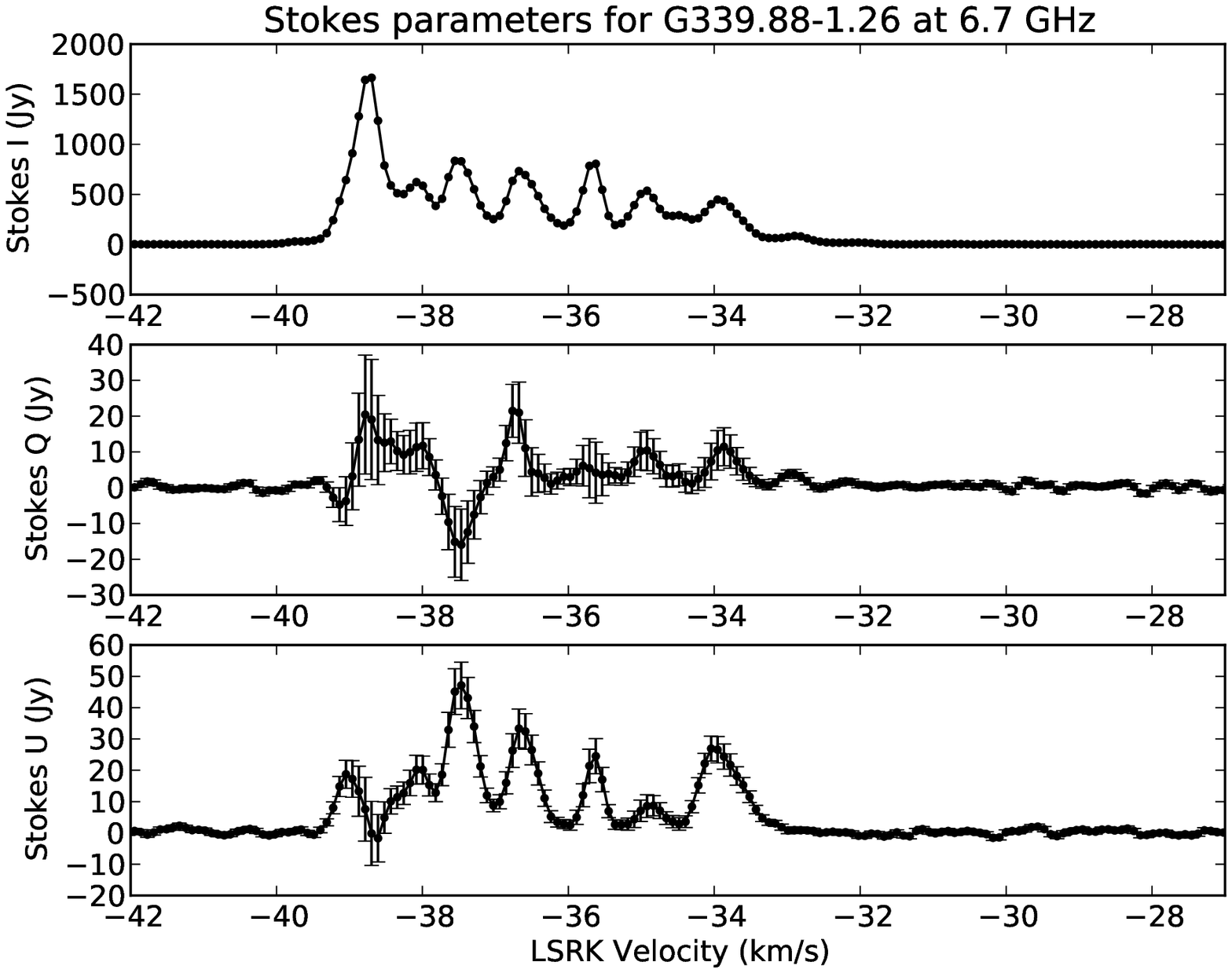}
&
\includegraphics[scale = 0.35]{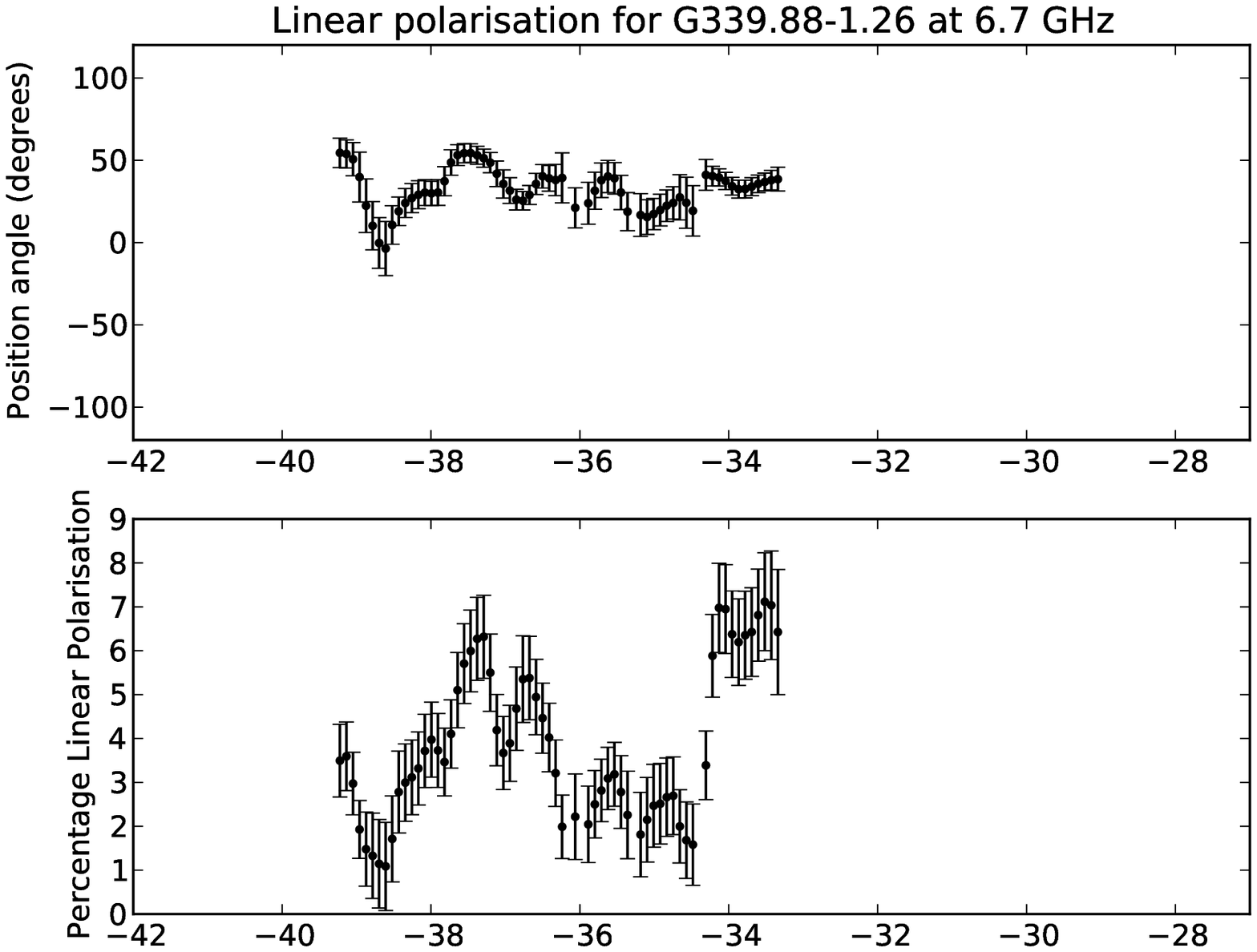}
\\
\begin{sideways}
2011 February 12
\end{sideways}
\includegraphics[scale = 0.35]{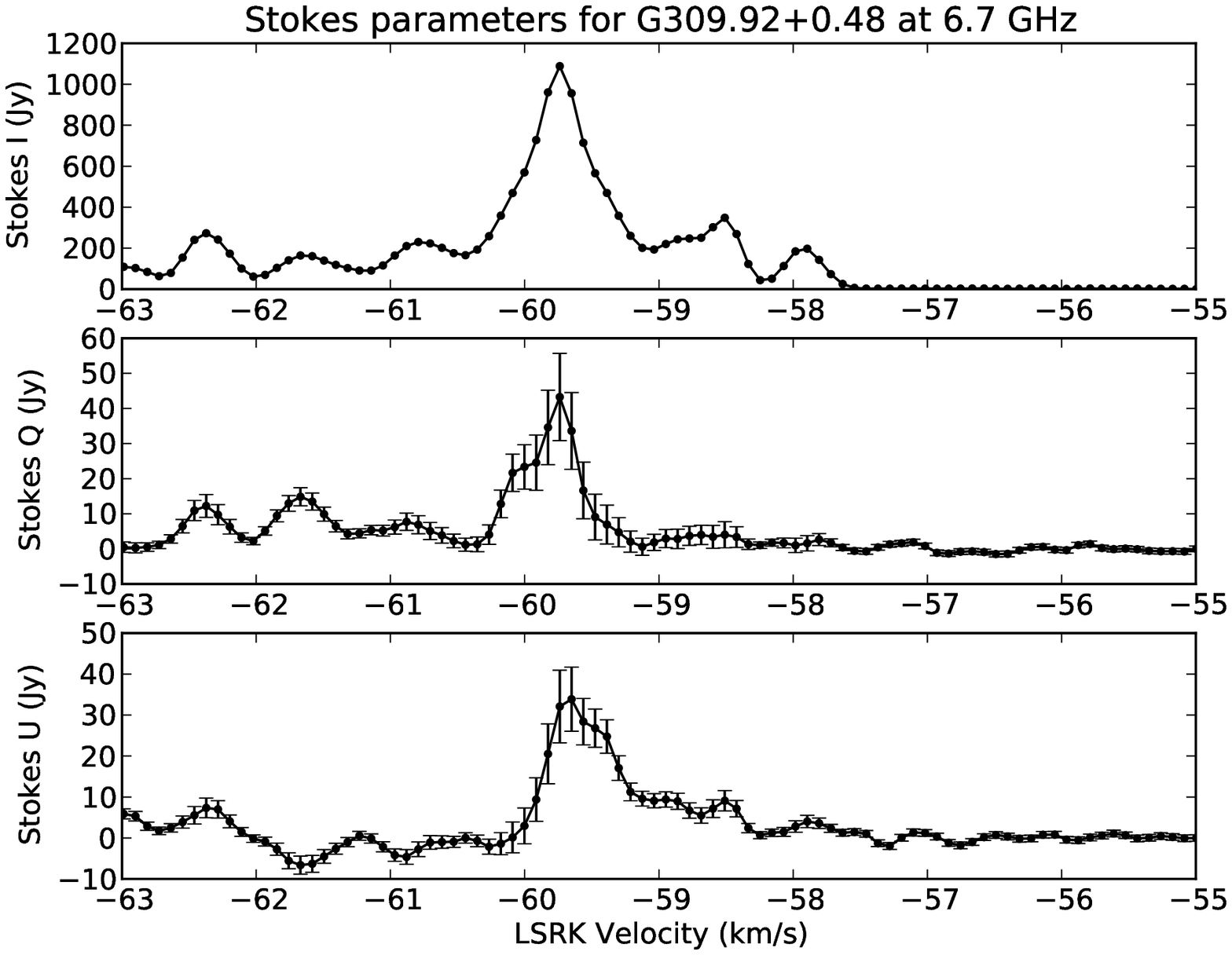}
&
\includegraphics[scale = 0.35]{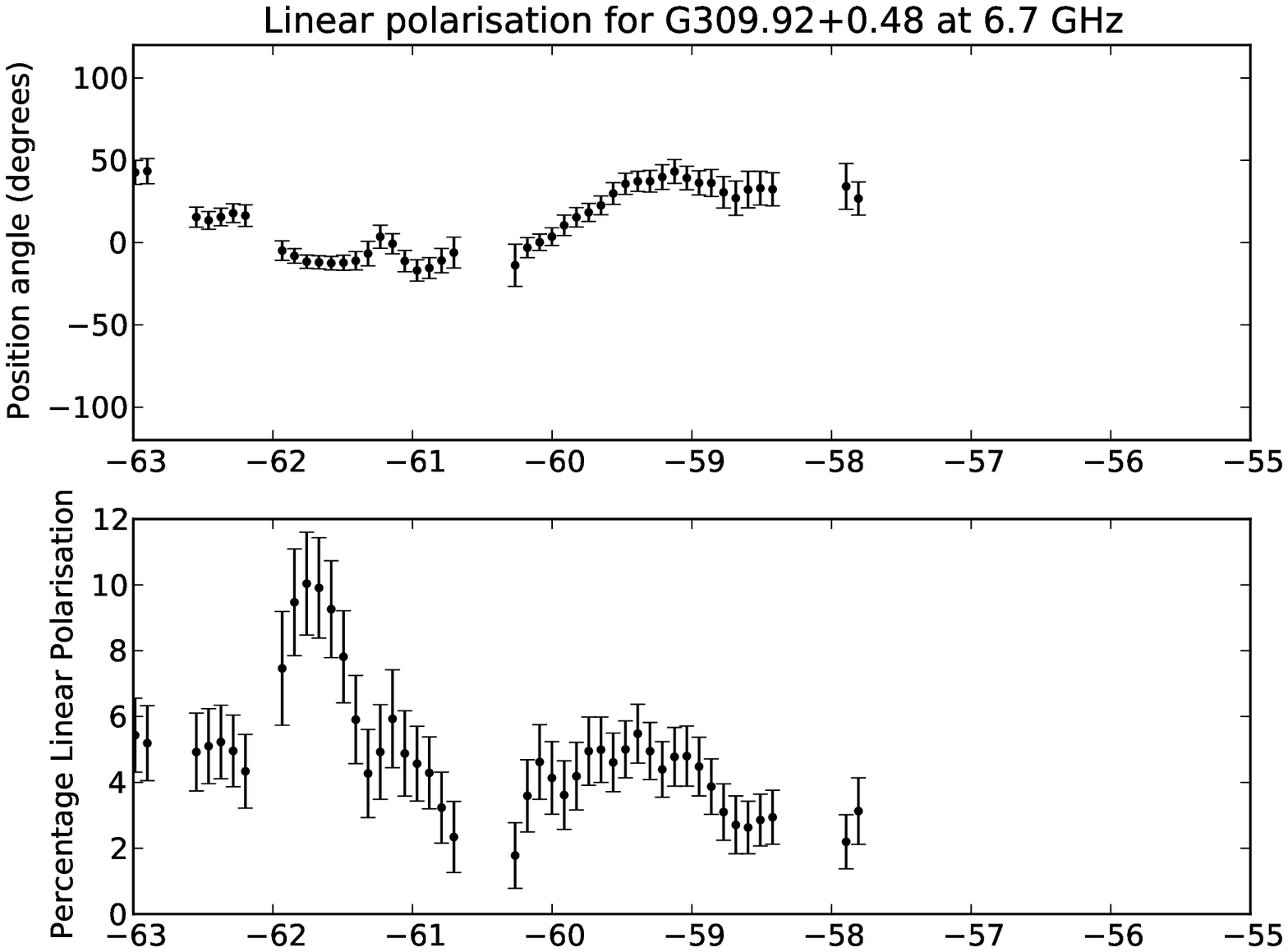}
\\
\begin{sideways}
2010 September 28-30
\end{sideways}
\includegraphics[scale = 0.35]{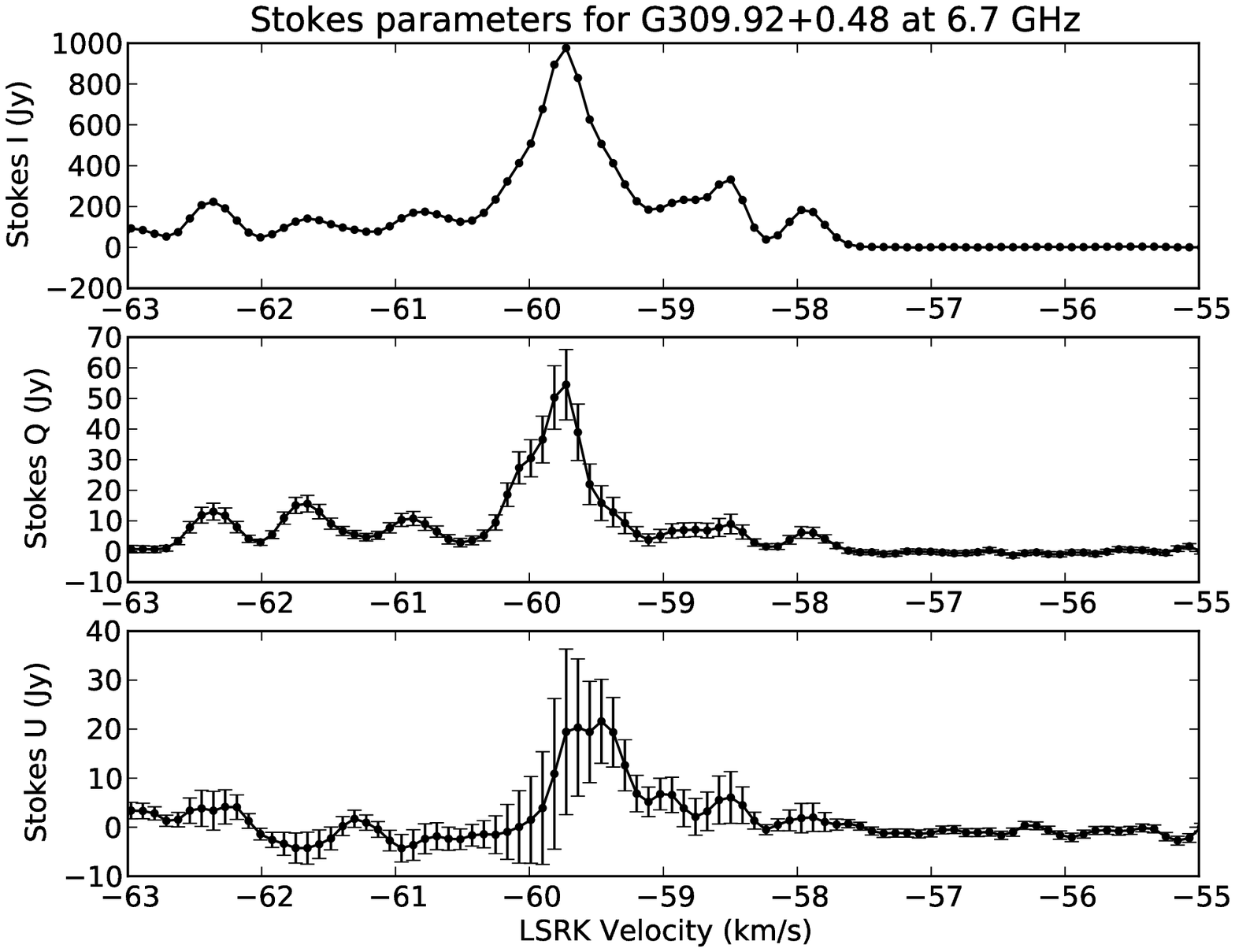}
&
\includegraphics[scale = 0.35]{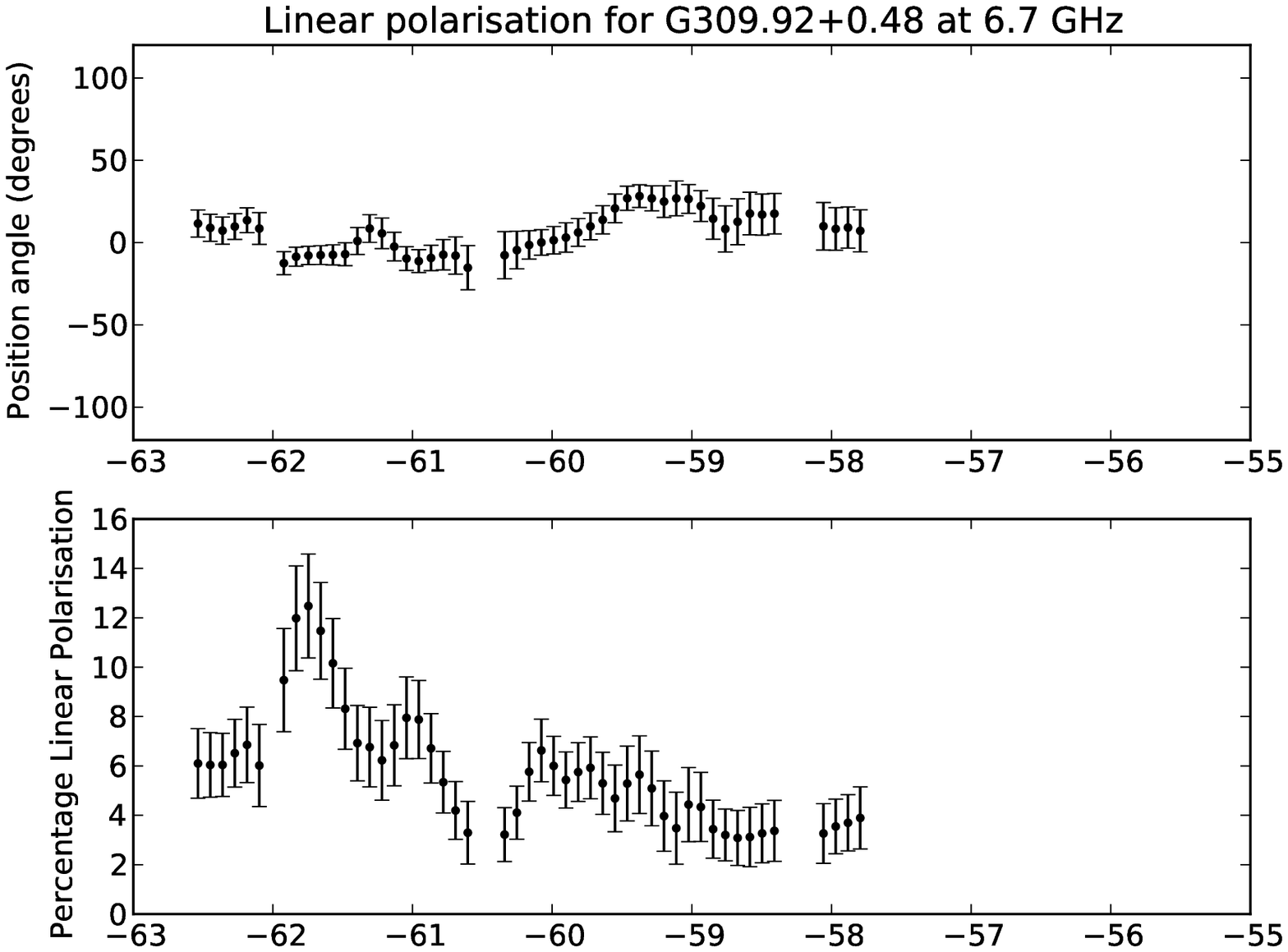}
\end{tabular}

\clearpage

\begin{tabular}{cc}
\begin{sideways}
2011 February 12
\end{sideways}
\includegraphics[scale = 0.35]{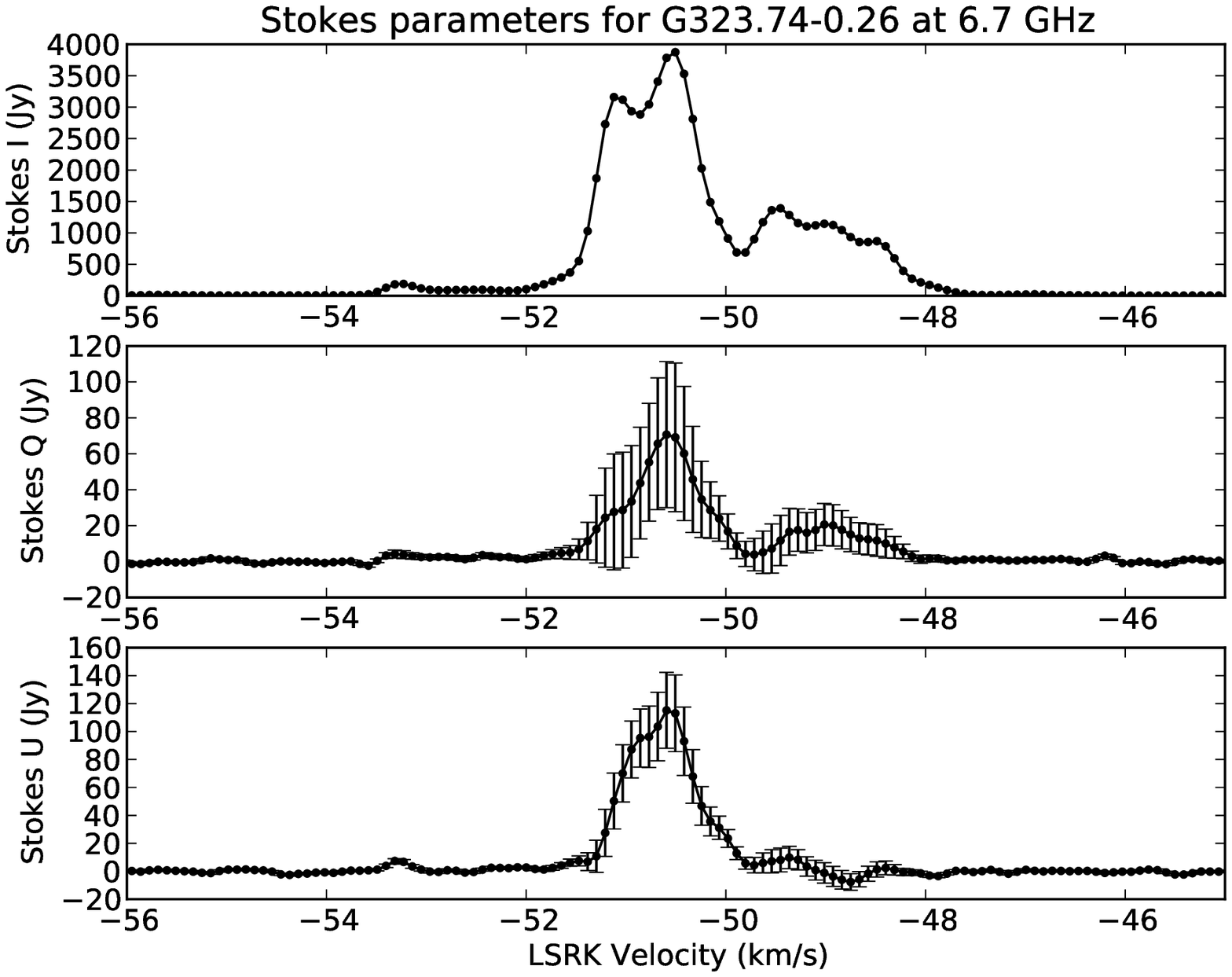}
&
\includegraphics[scale = 0.35]{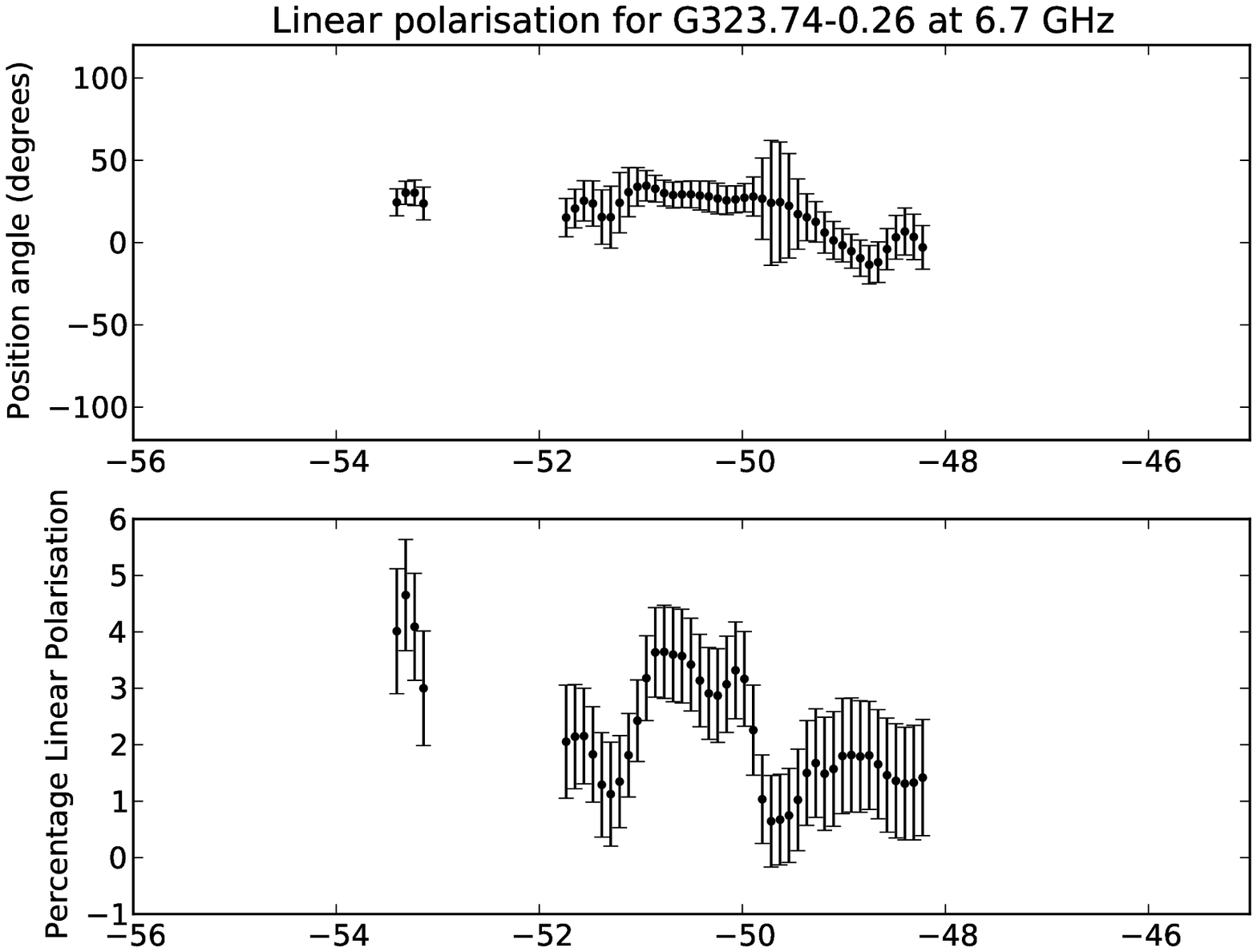}
\\
\begin{sideways}
2010 September 28-30
\end{sideways}
\includegraphics[scale = 0.35]{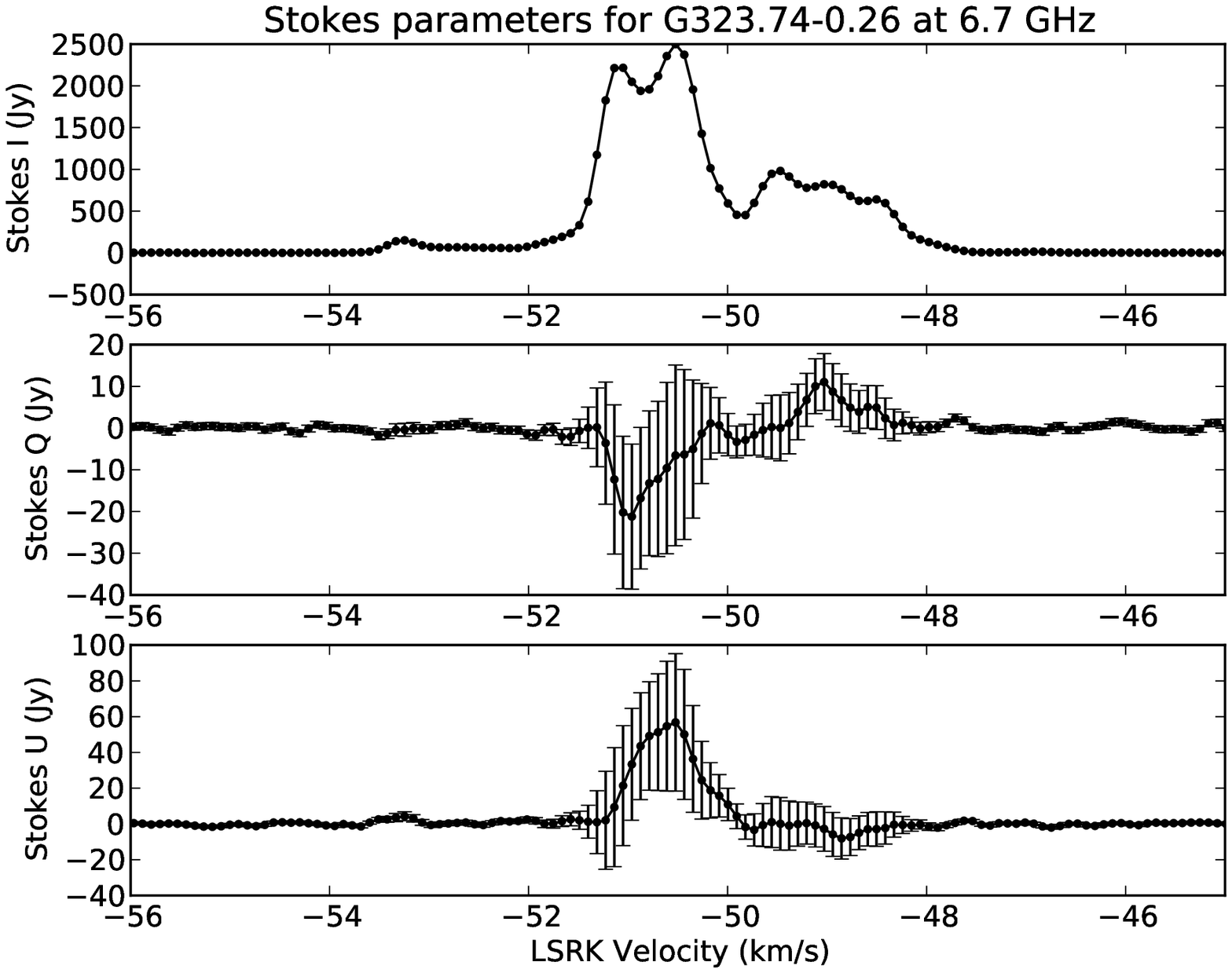}
&
\includegraphics[scale = 0.35]{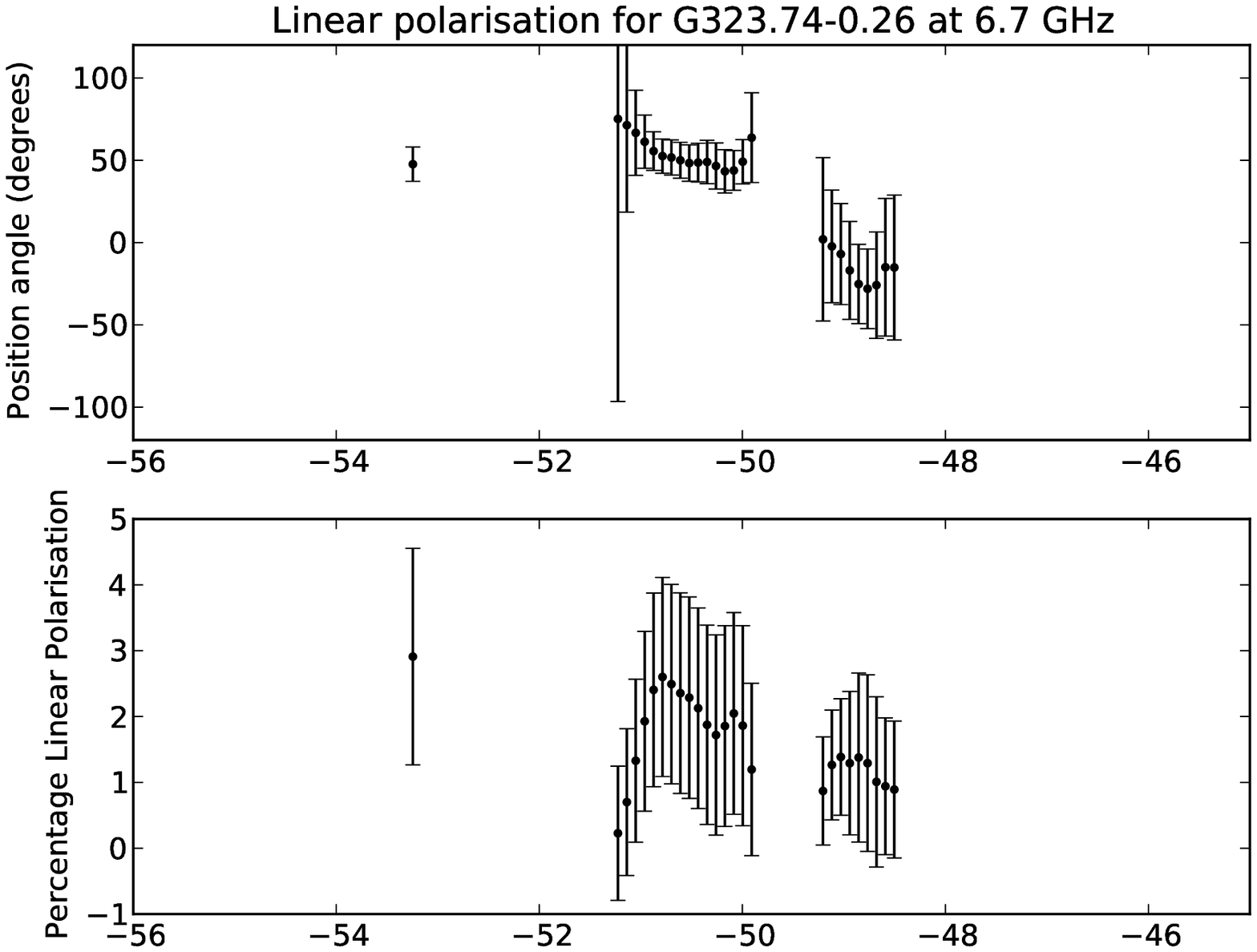}
\\
\begin{sideways}
2011 February 12
\end{sideways}
\includegraphics[scale = 0.35]{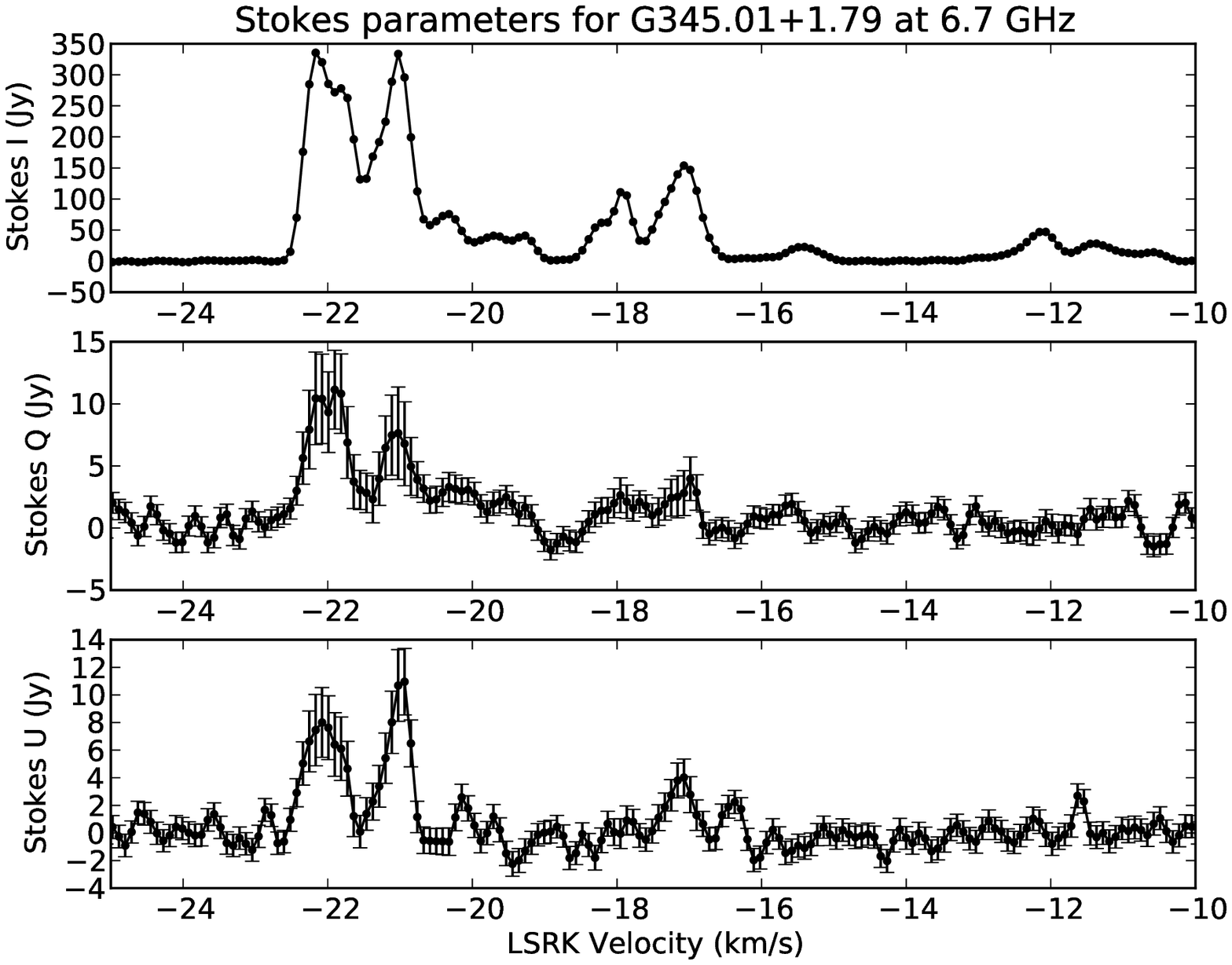}
&
\includegraphics[scale = 0.35]{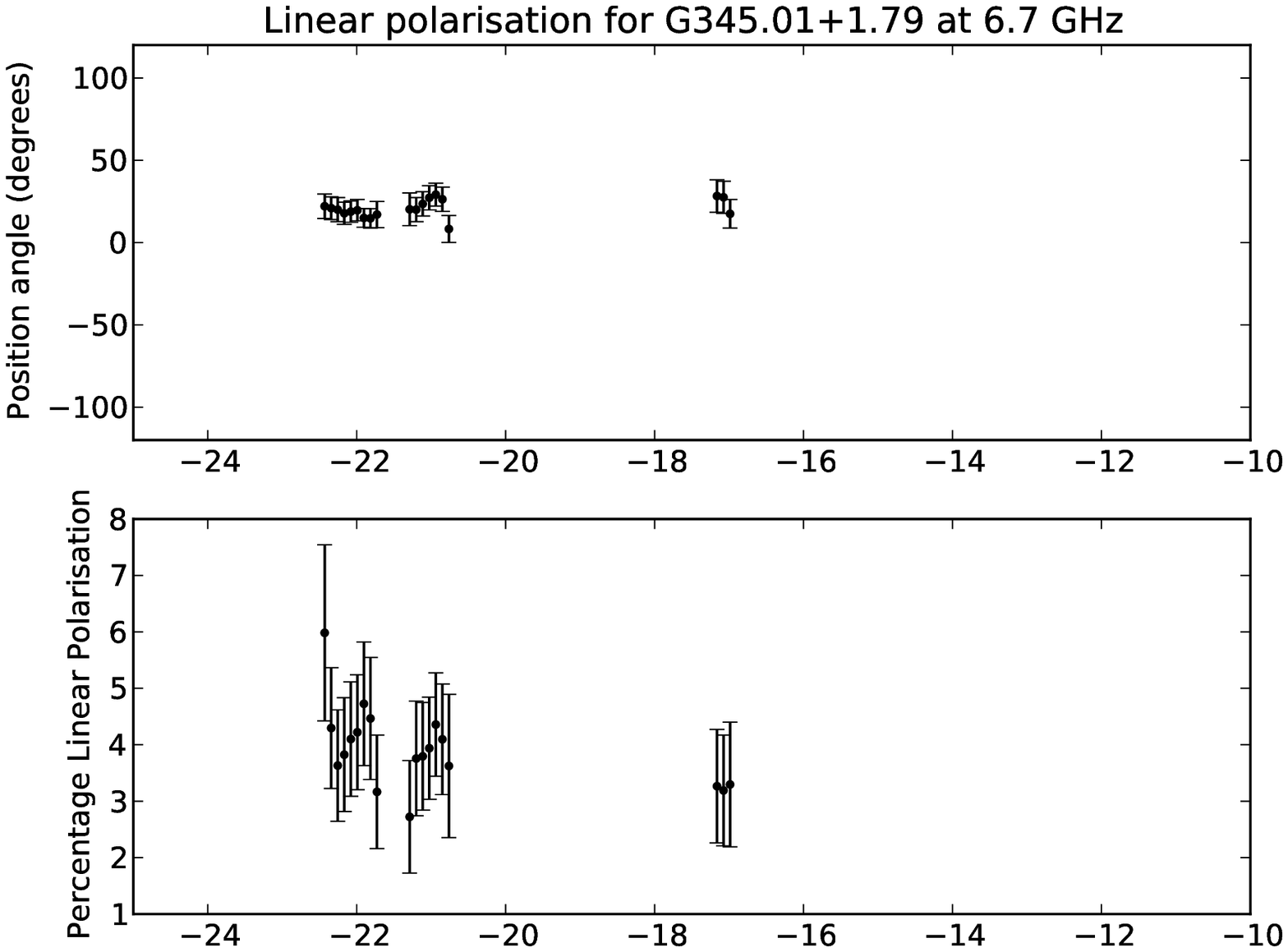}
\\
\begin{sideways}
2010 September 28-30
\end{sideways}
\includegraphics[scale = 0.35]{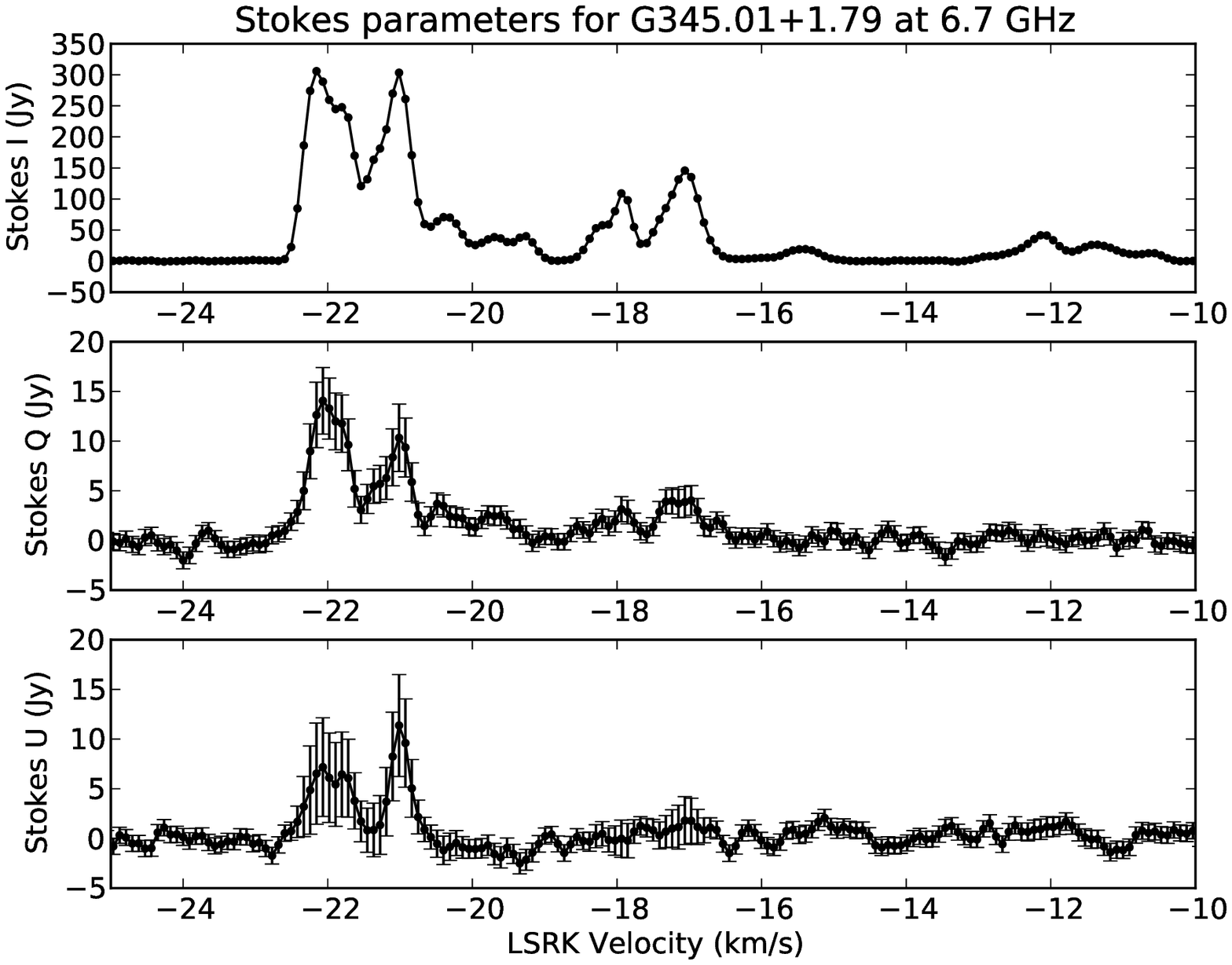}
&
\includegraphics[scale = 0.35]{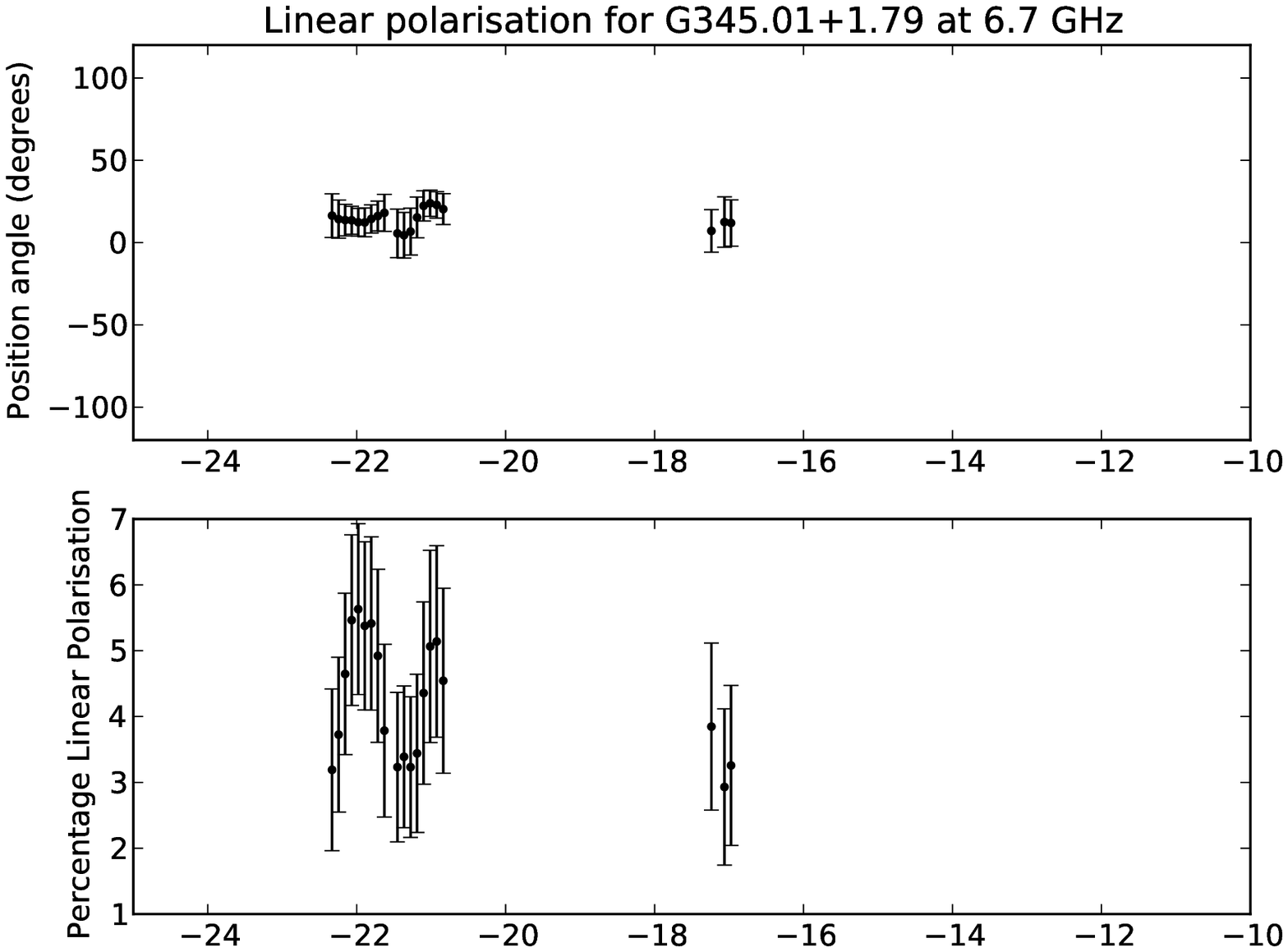}
\end{tabular}

\clearpage

\begin{tabular}{cc}
\begin{sideways}
2011 February 12
\end{sideways}
\includegraphics[scale = 0.35]{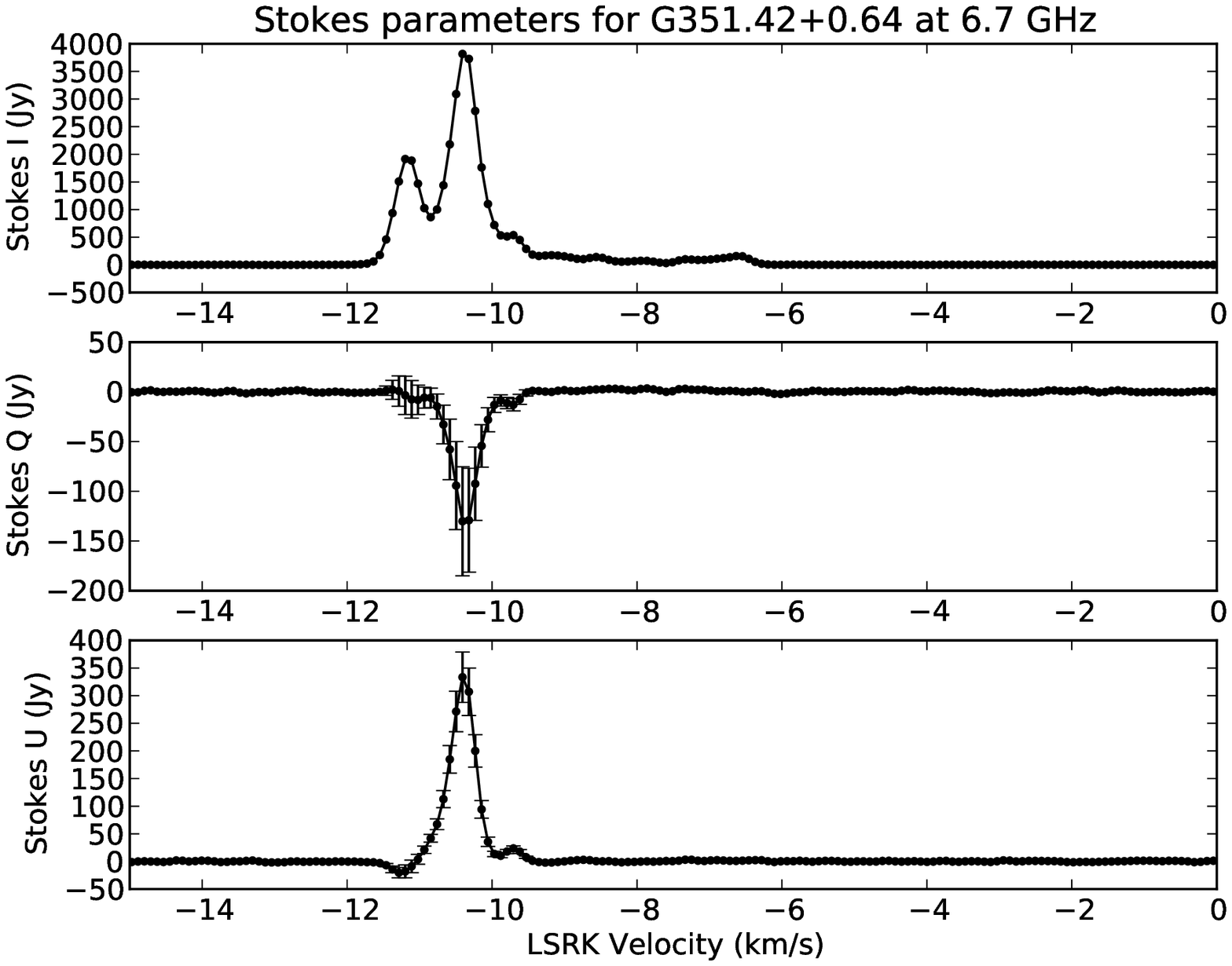}
&
\includegraphics[scale = 0.35]{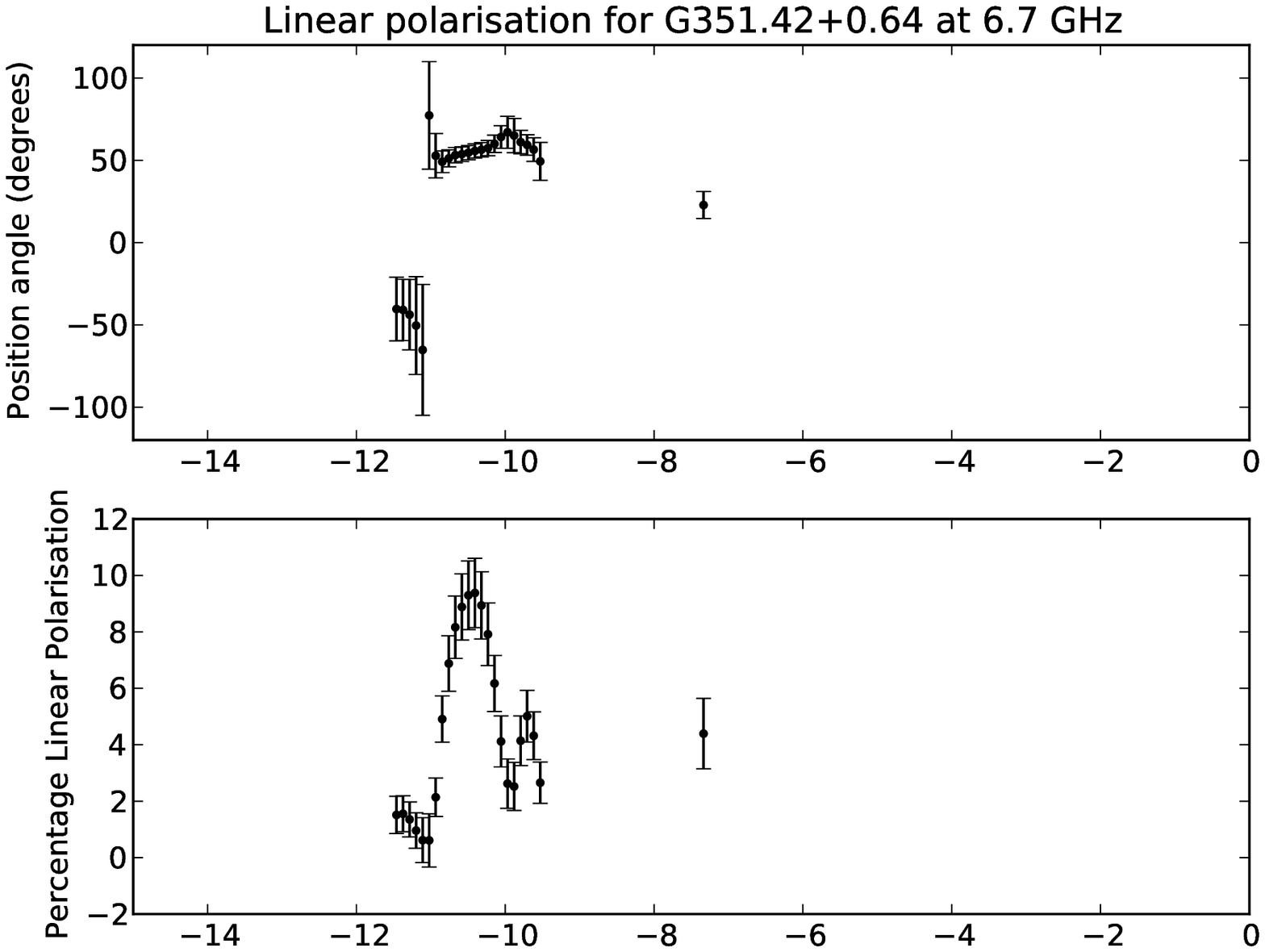}
\\
\begin{sideways}
2010 September 28-30
\end{sideways}
\includegraphics[scale = 0.35]{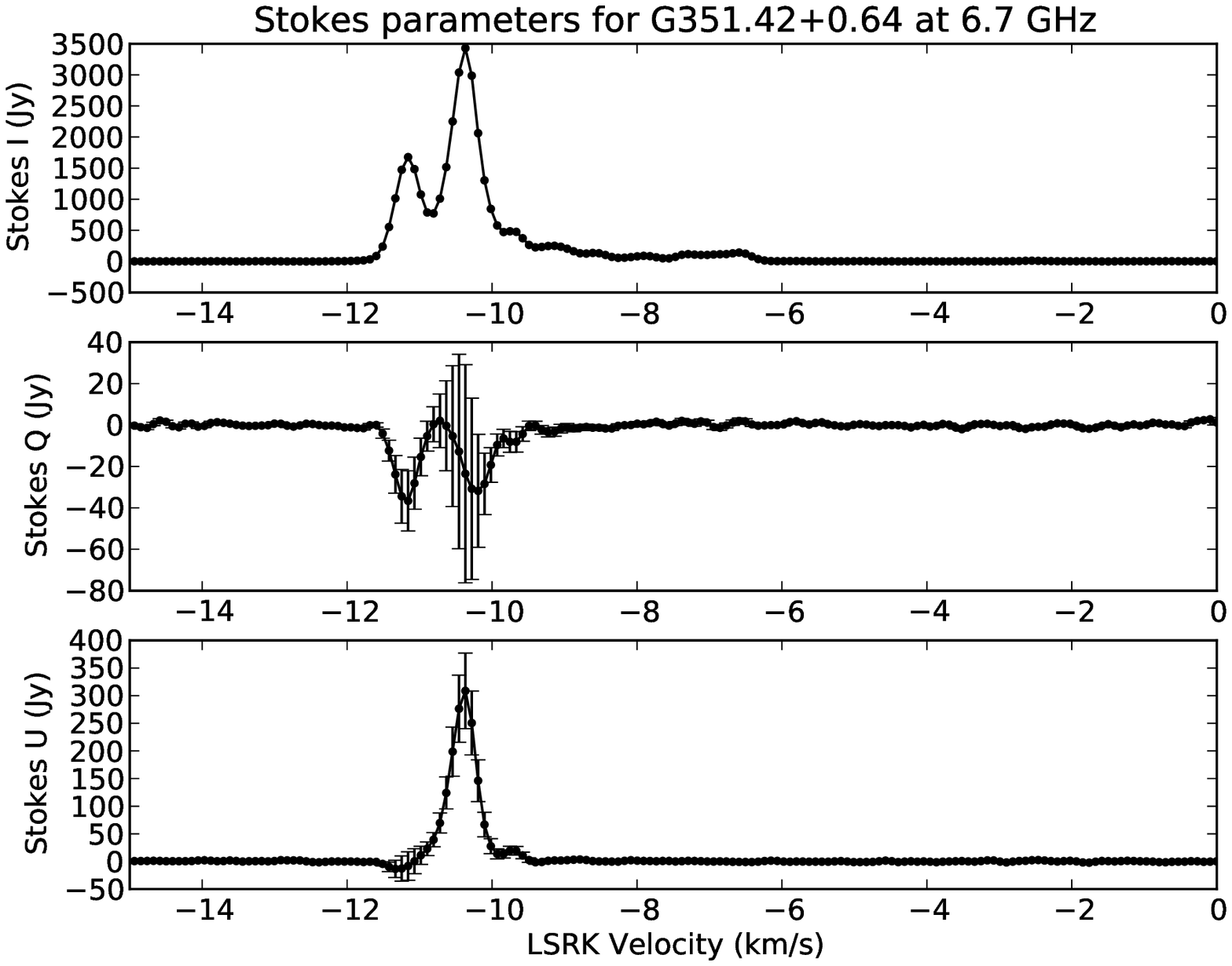}
&
\includegraphics[scale = 0.35]{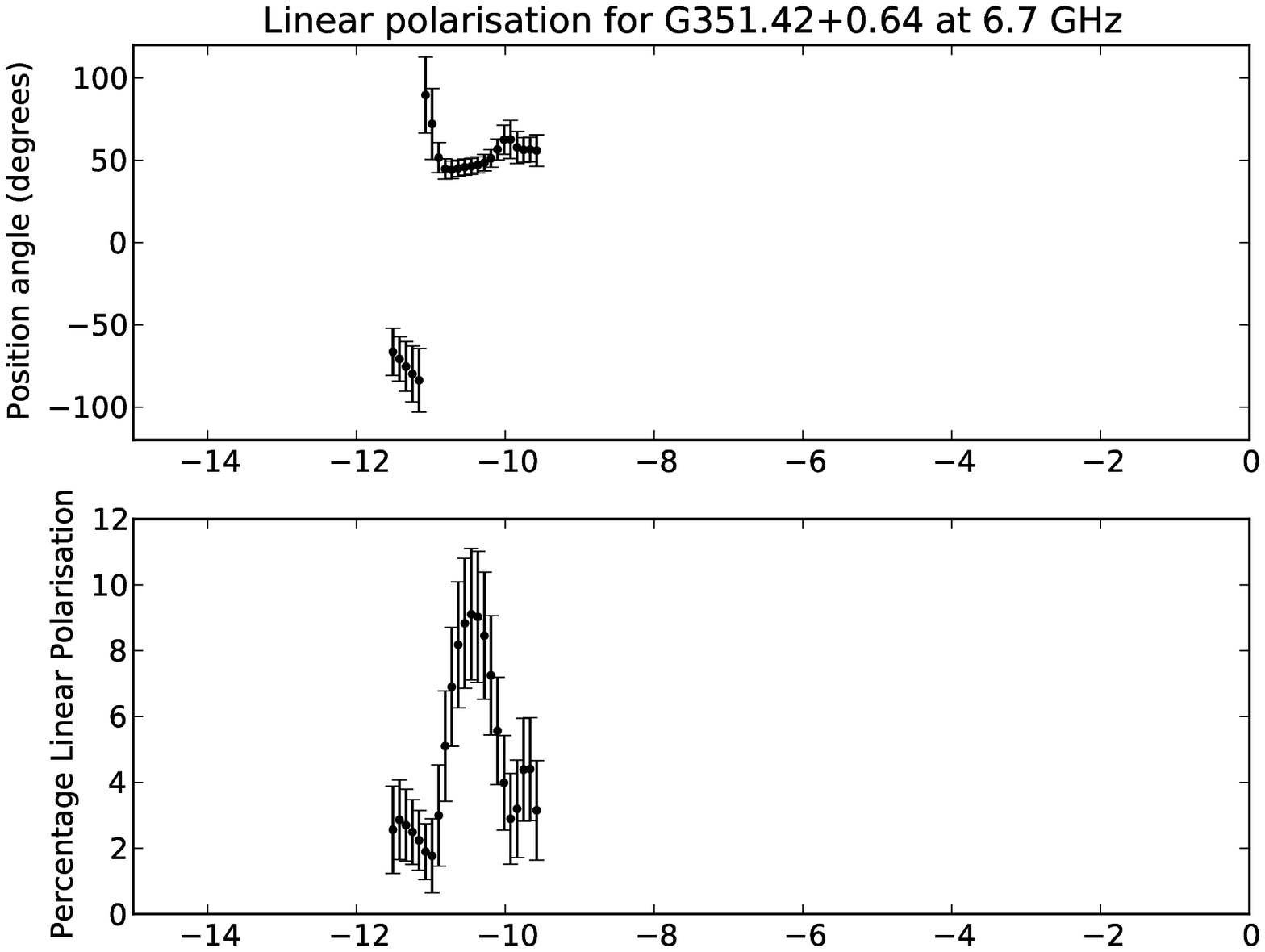}
\\
\begin{sideways}
2011 February 12
\end{sideways}
\includegraphics[scale = 0.35]{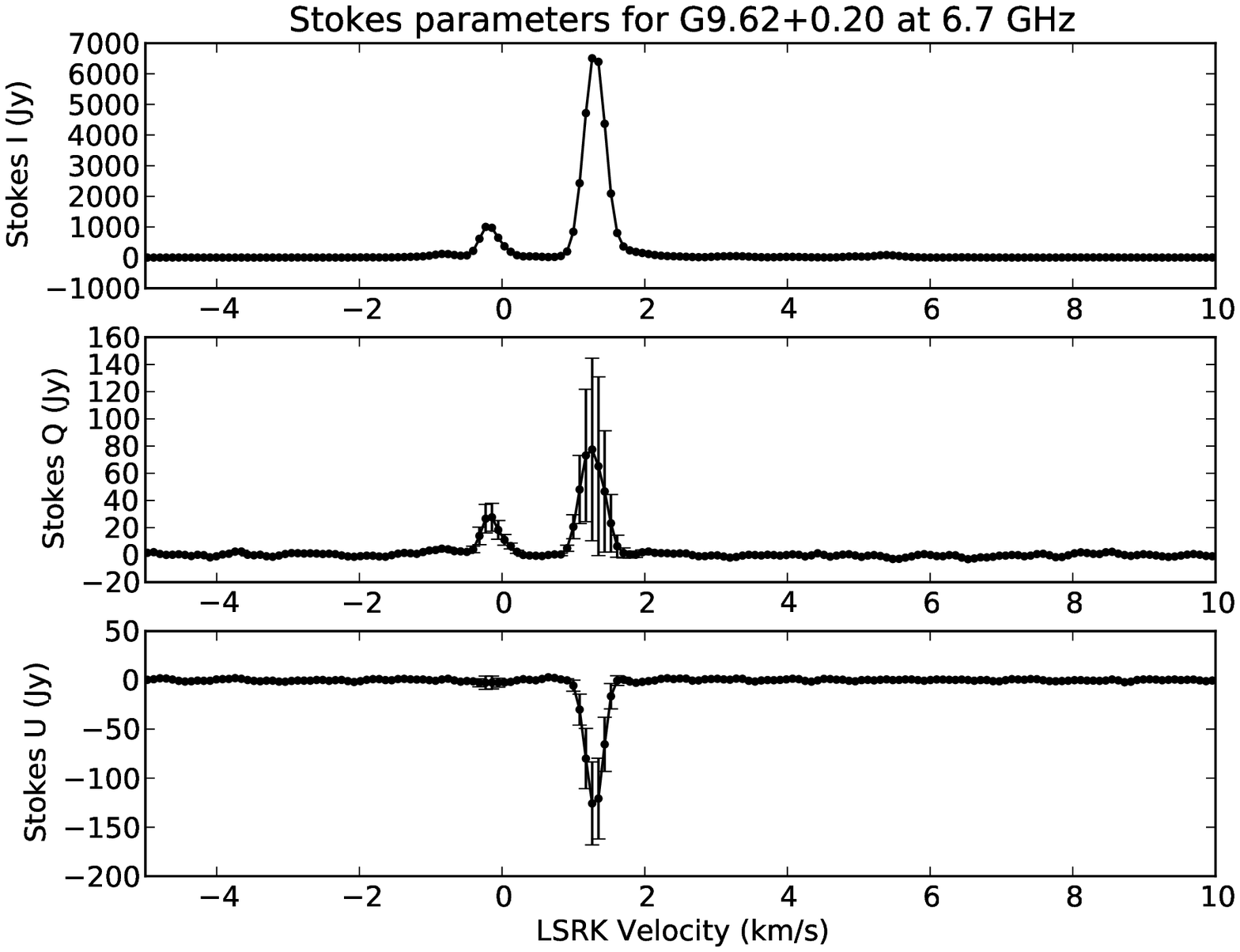}
&
\includegraphics[scale = 0.35]{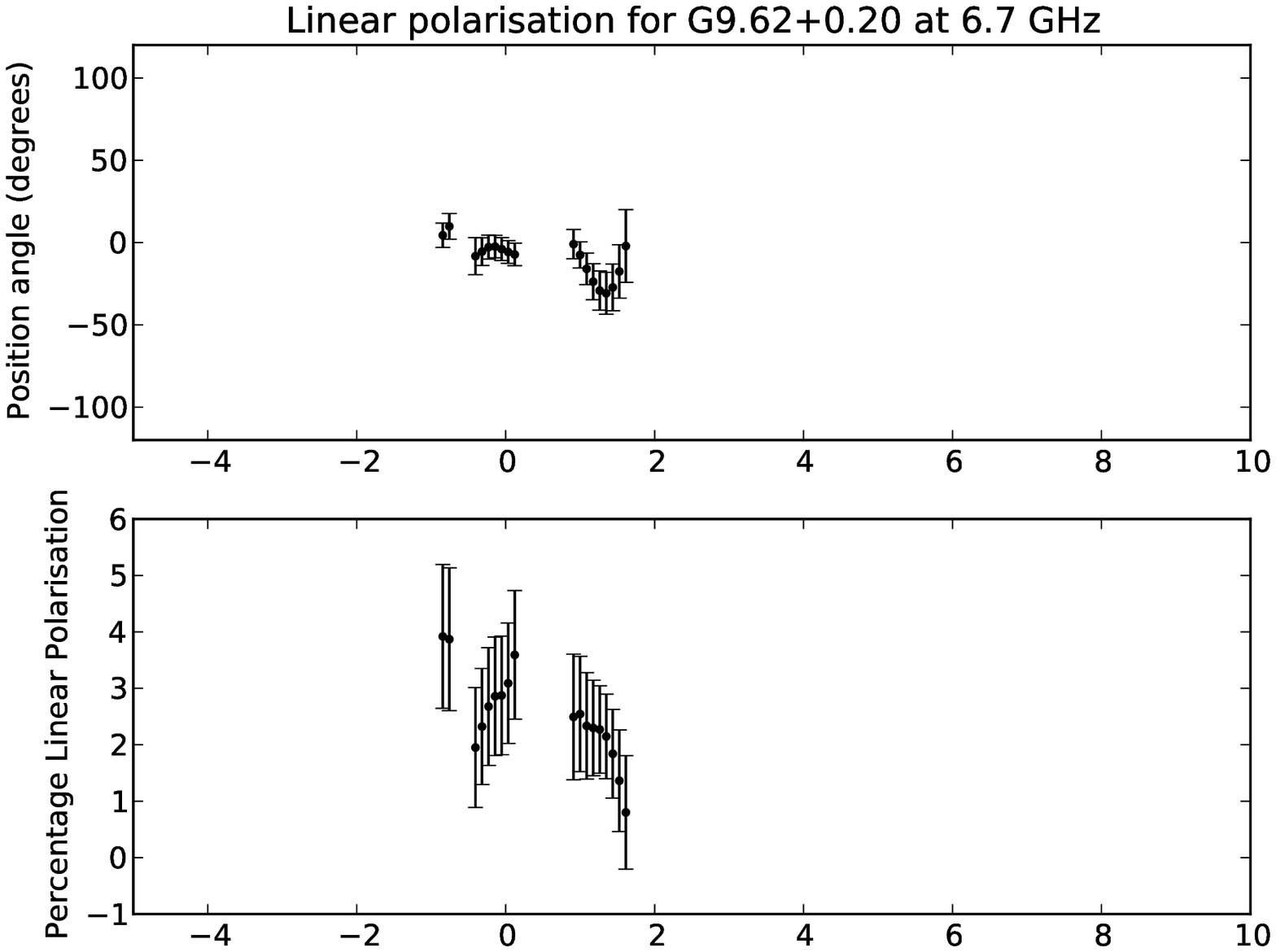}
\\
\begin{sideways}
2010 September 28-30
\end{sideways}
\includegraphics[scale = 0.35]{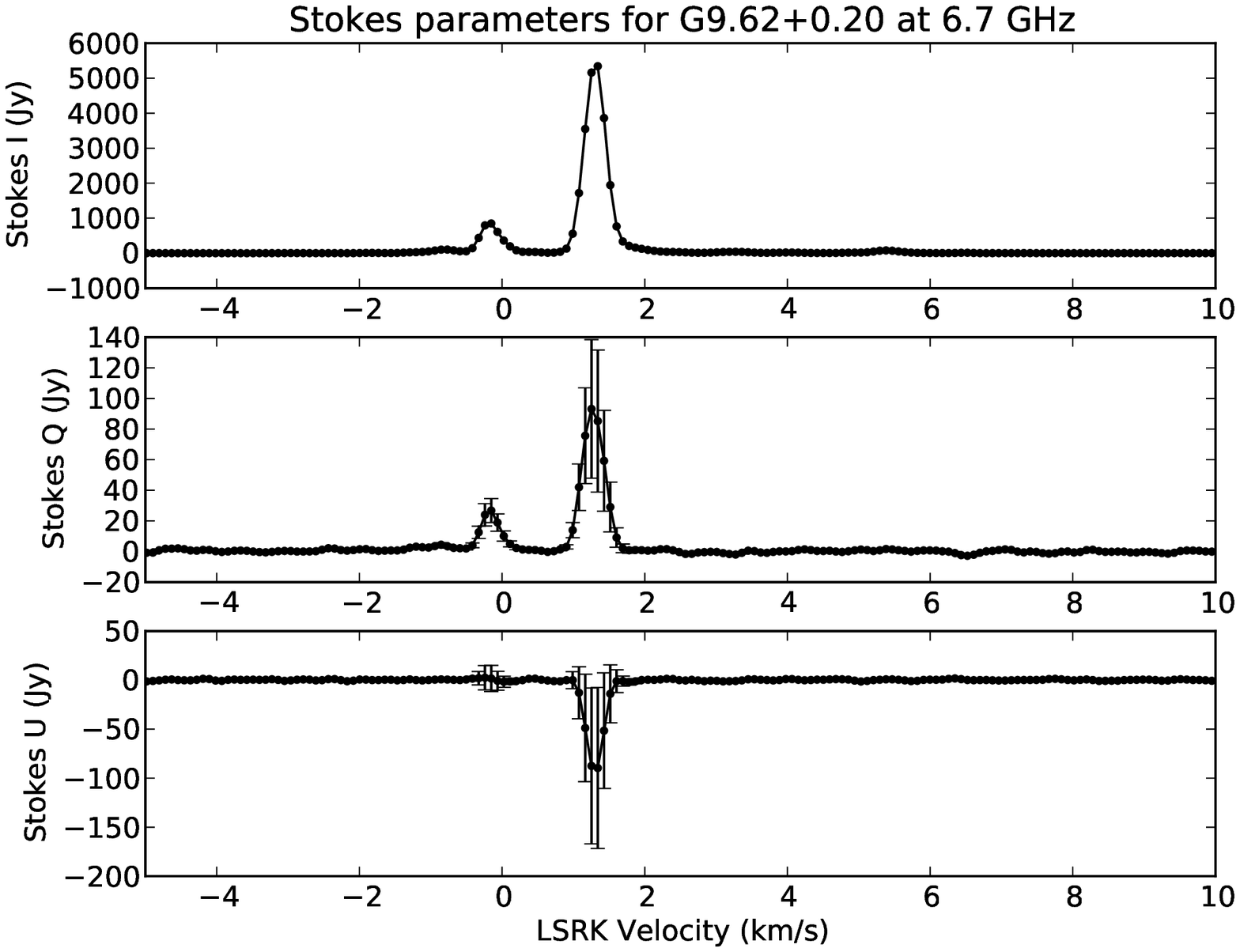}
&
\includegraphics[scale = 0.35]{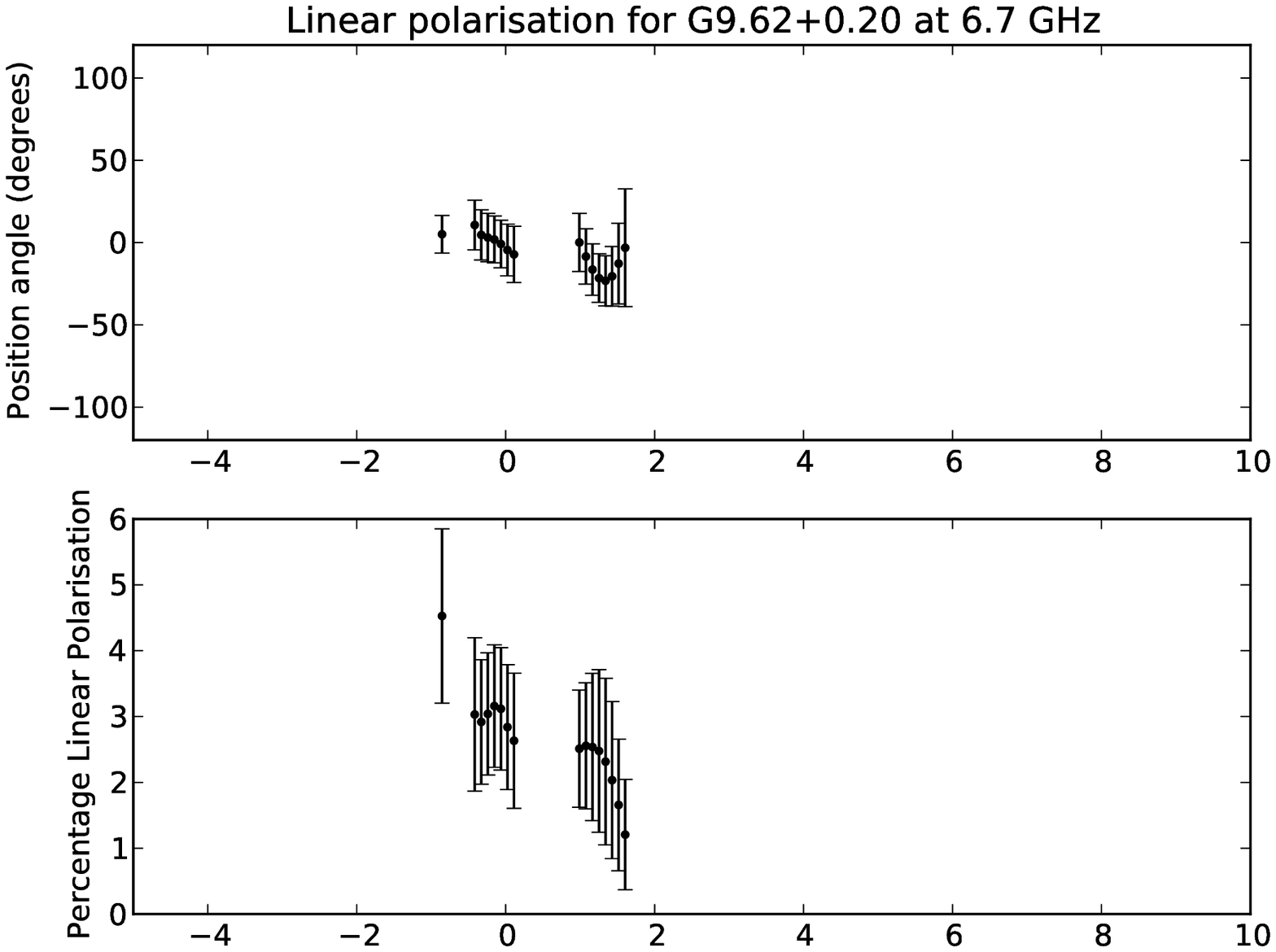}
\end{tabular}

\clearpage

\begin{minipage}{\textwidth}
\subsection*{Sources with significant linear polarisation}
These sources were observed 2010 September 28-30. Linear polarisation above the 5 times RMS level was considered to be a detection, plots of the position angle and fractional linear polarisation plots are restricted to those spectral channels which meet this criterion.
\end{minipage}

\begin{tabular}{cc}
\includegraphics[scale = 0.35]{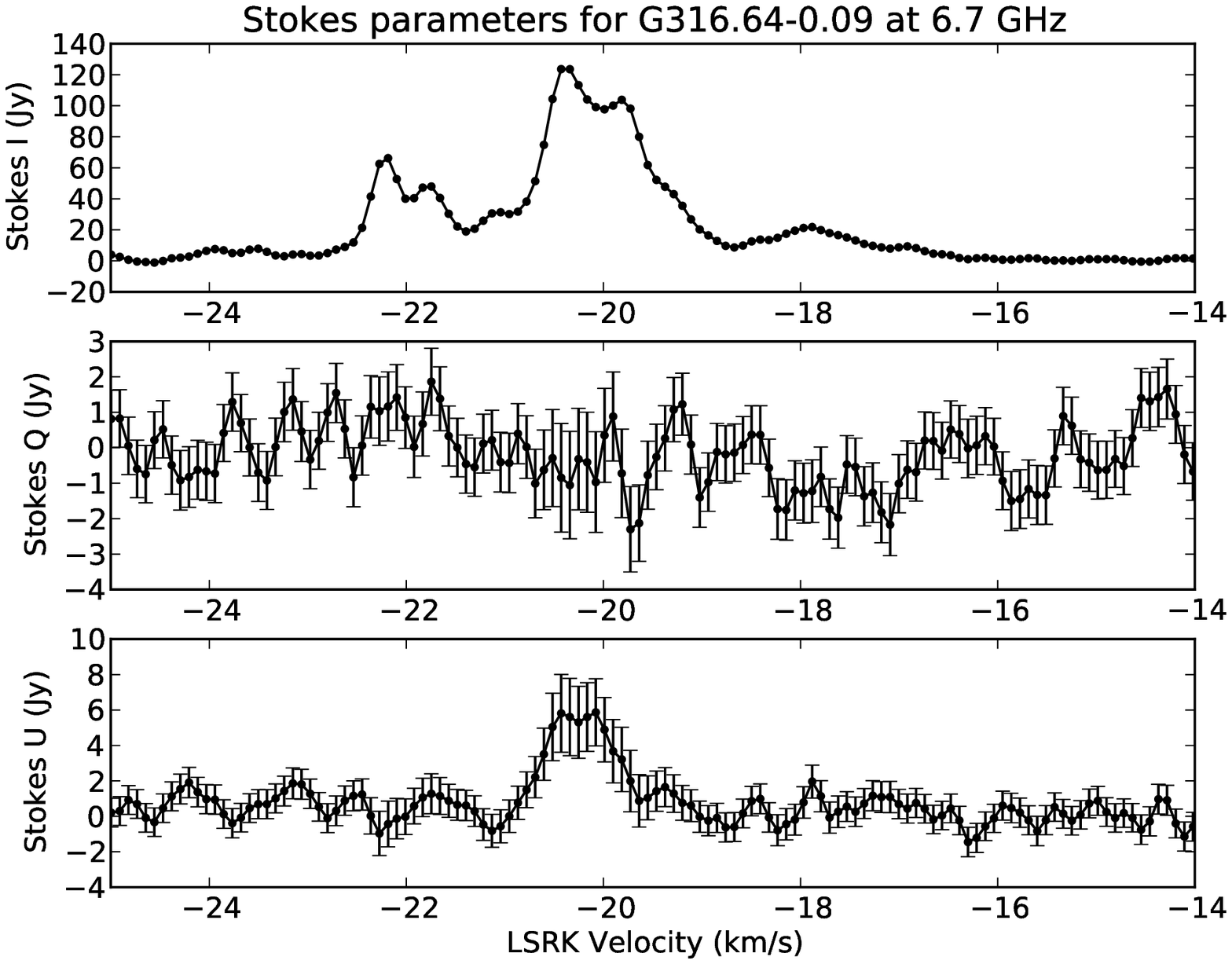}
&
\includegraphics[scale = 0.35]{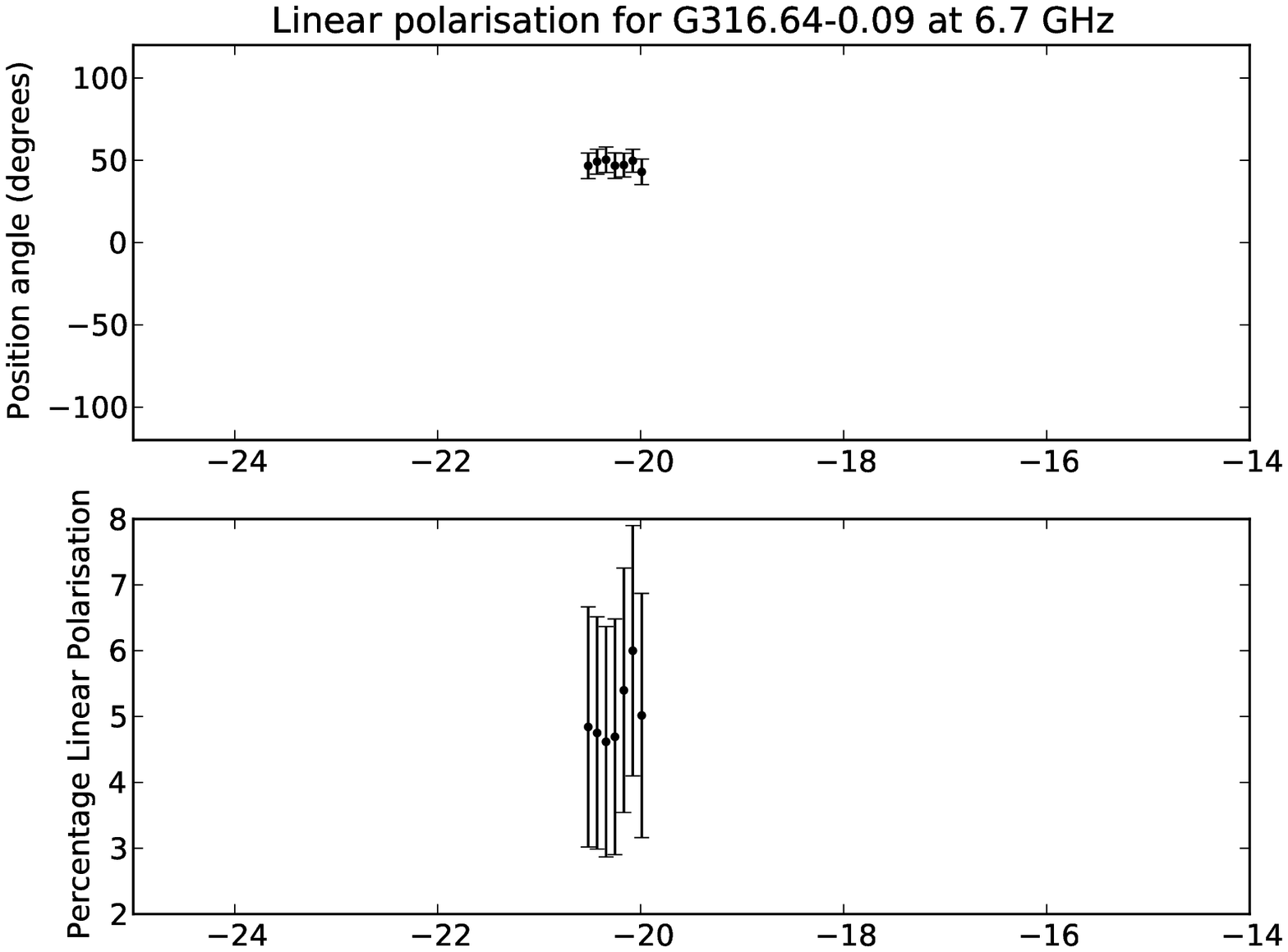}
\\
\includegraphics[scale = 0.35]{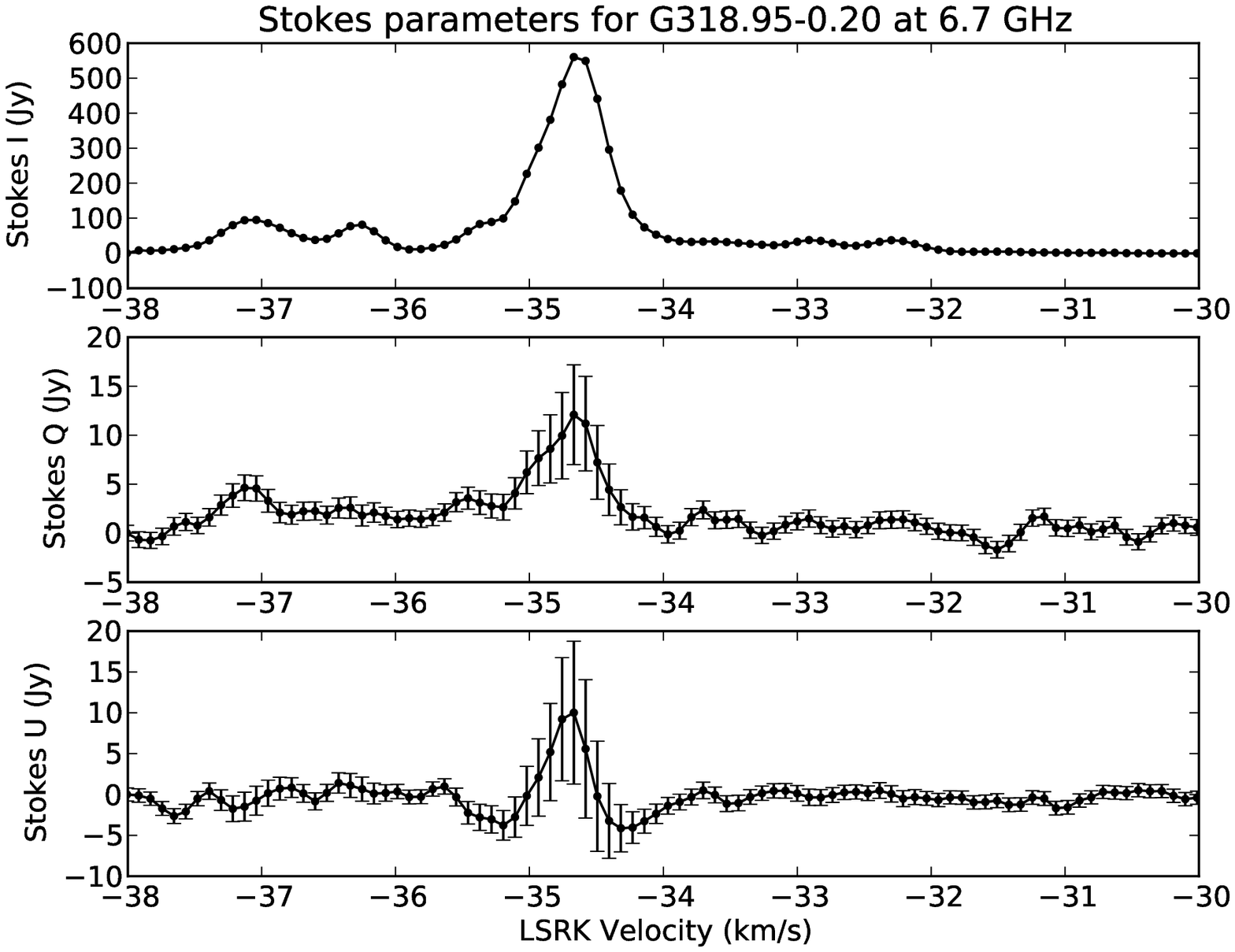}
&
\includegraphics[scale = 0.35]{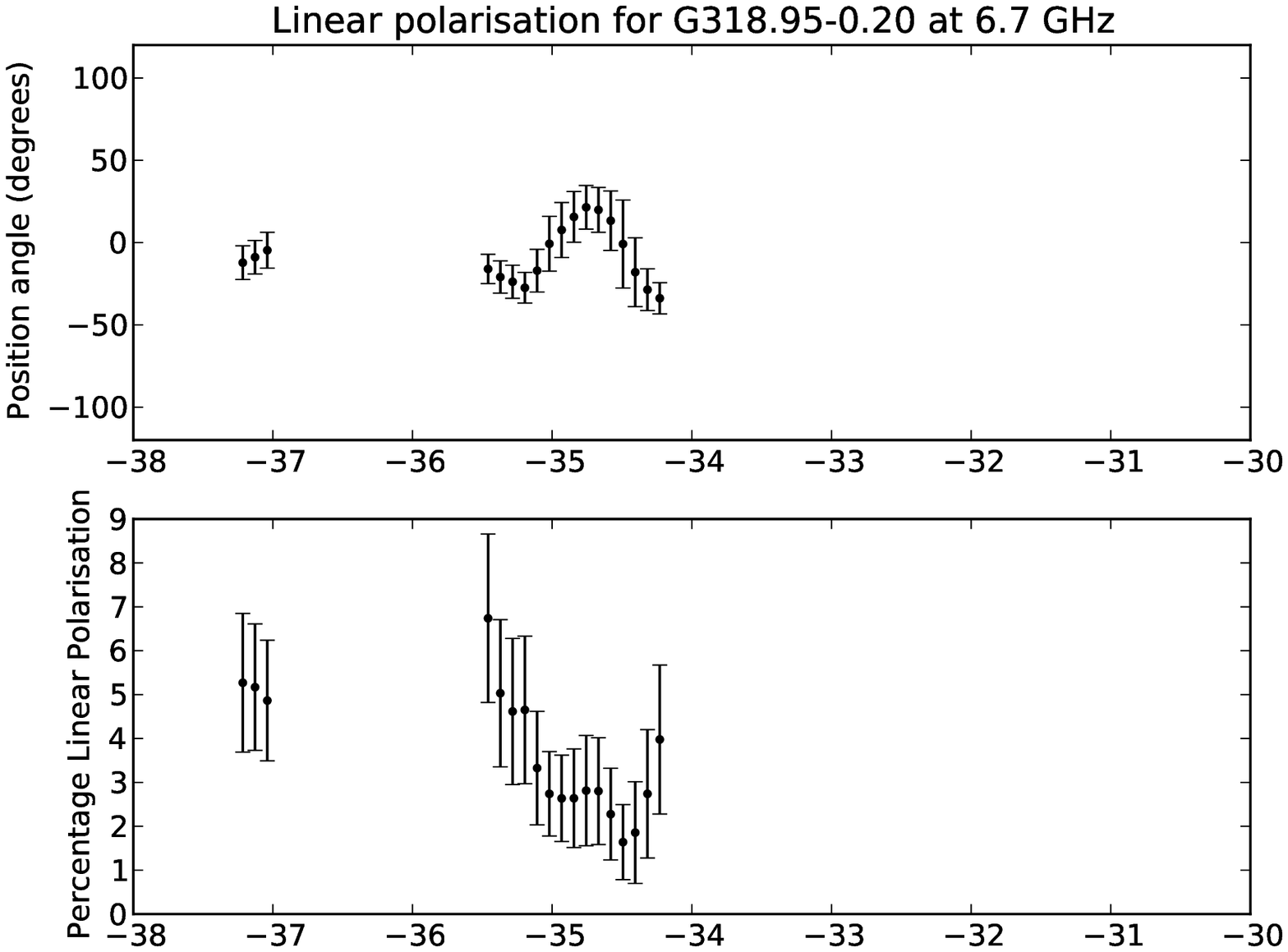}
\\
\includegraphics[scale = 0.35]{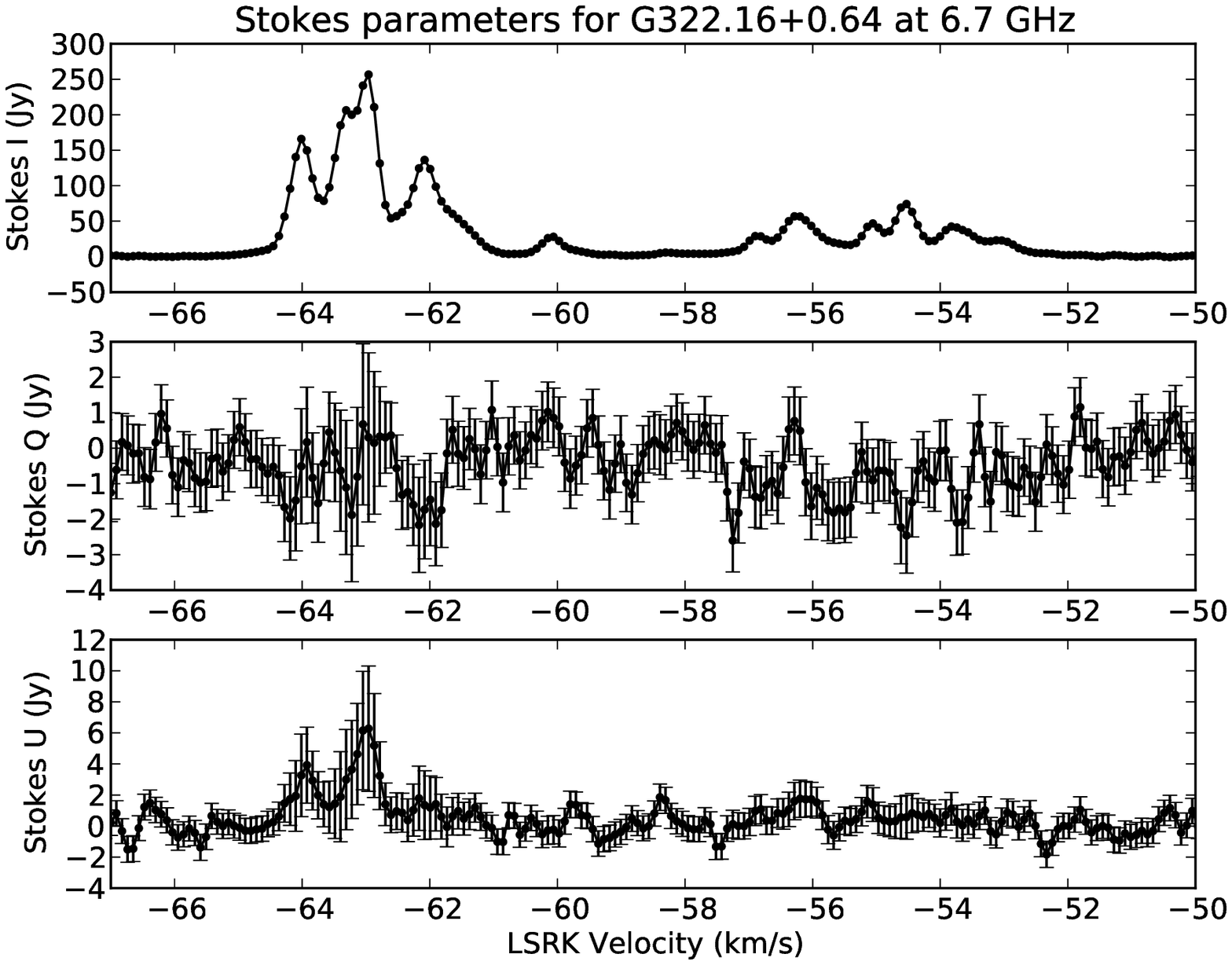}
&
\includegraphics[scale = 0.35]{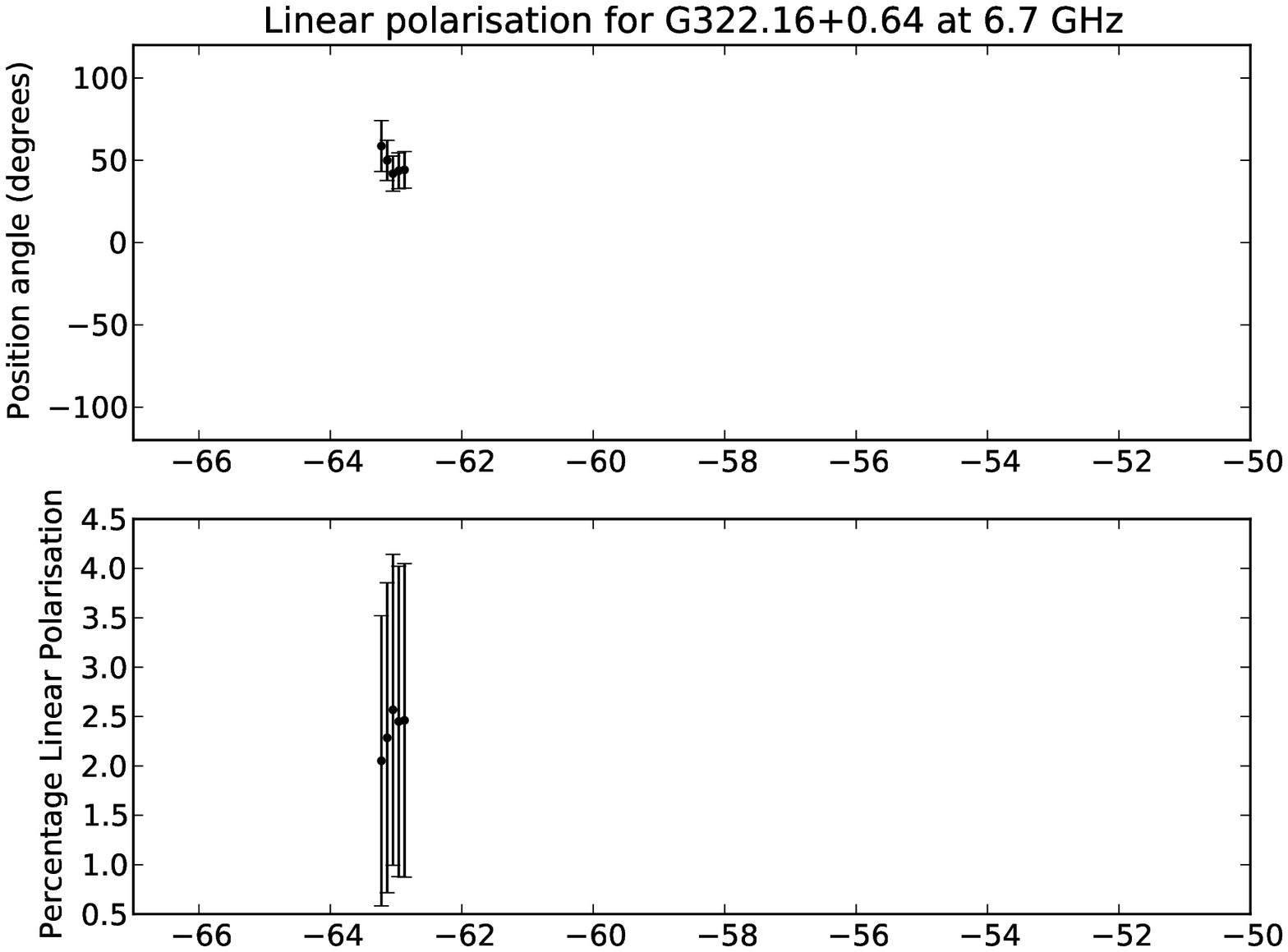}
\end{tabular}

\clearpage

\begin{tabular}{cc}
\includegraphics[scale = 0.35]{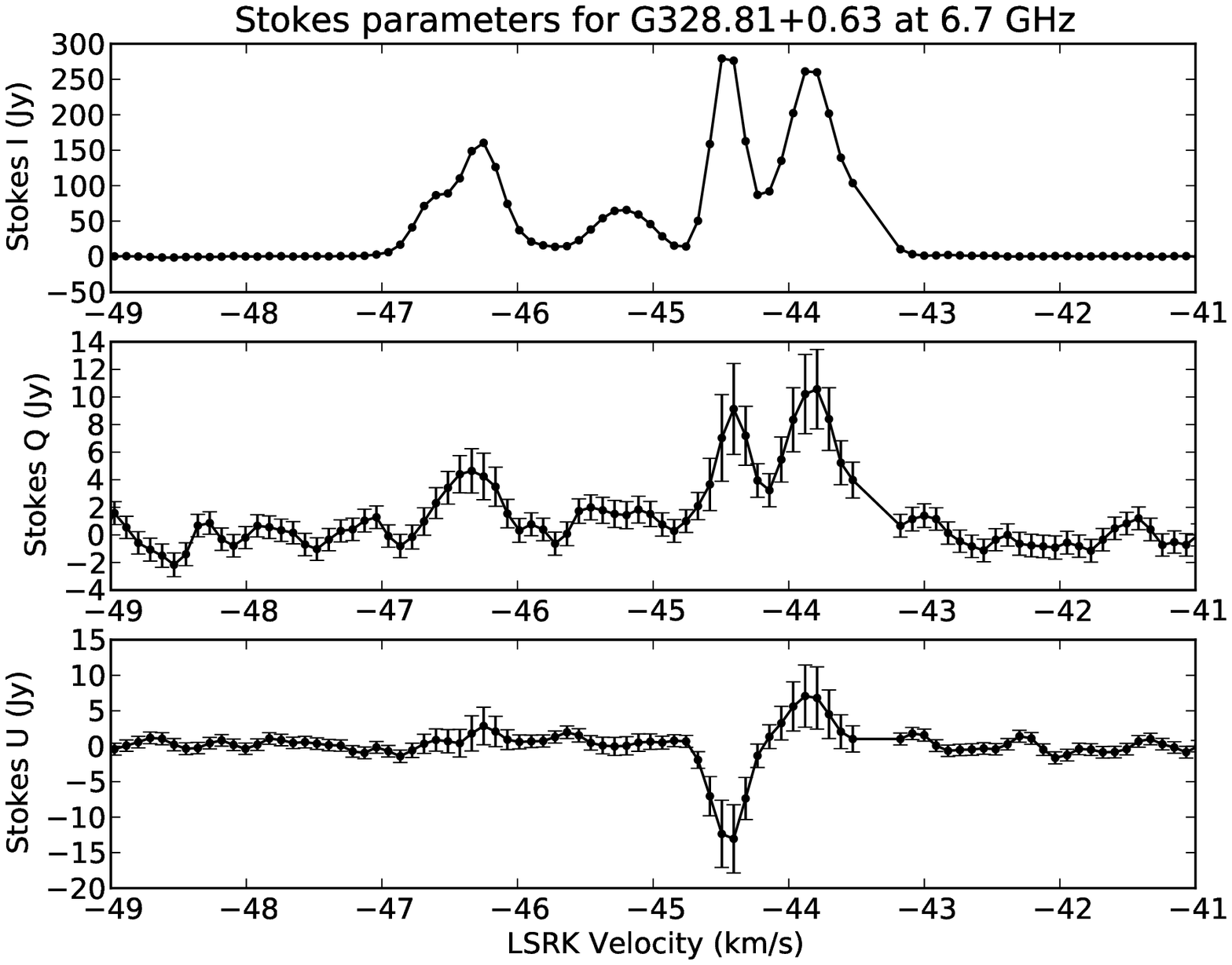}
&
\includegraphics[scale = 0.35]{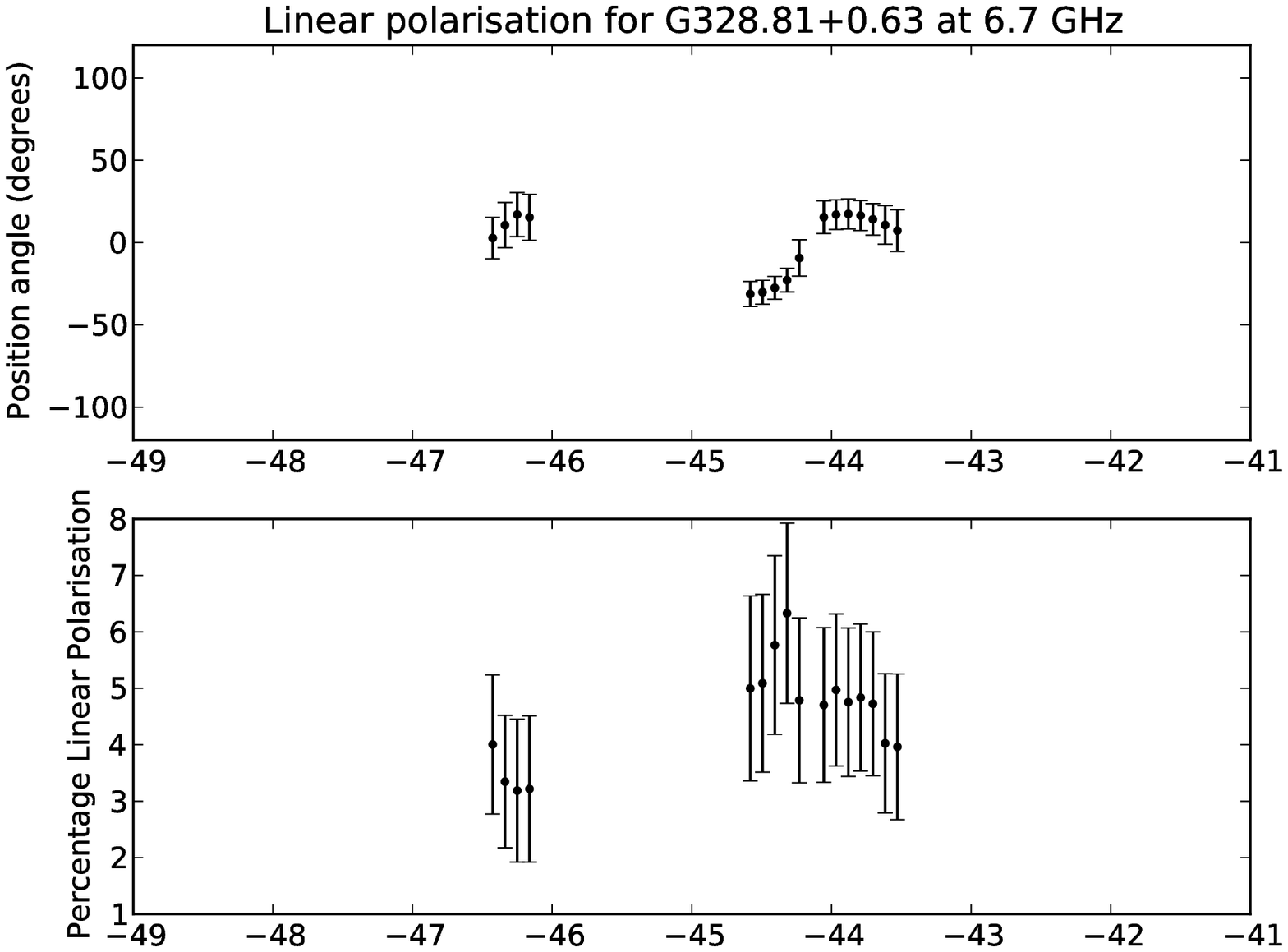}
\\
\includegraphics[scale = 0.35]{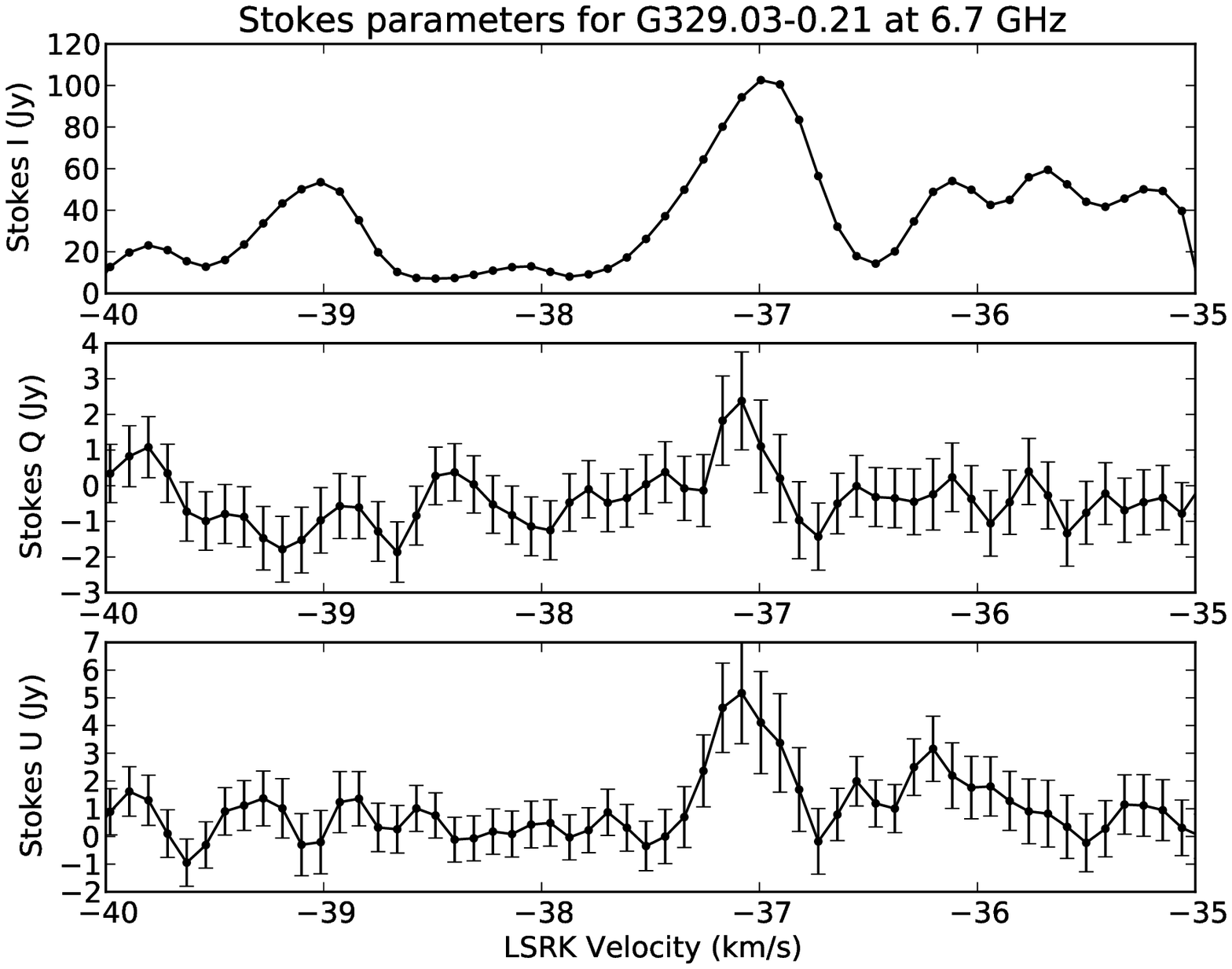}
&
\includegraphics[scale = 0.35]{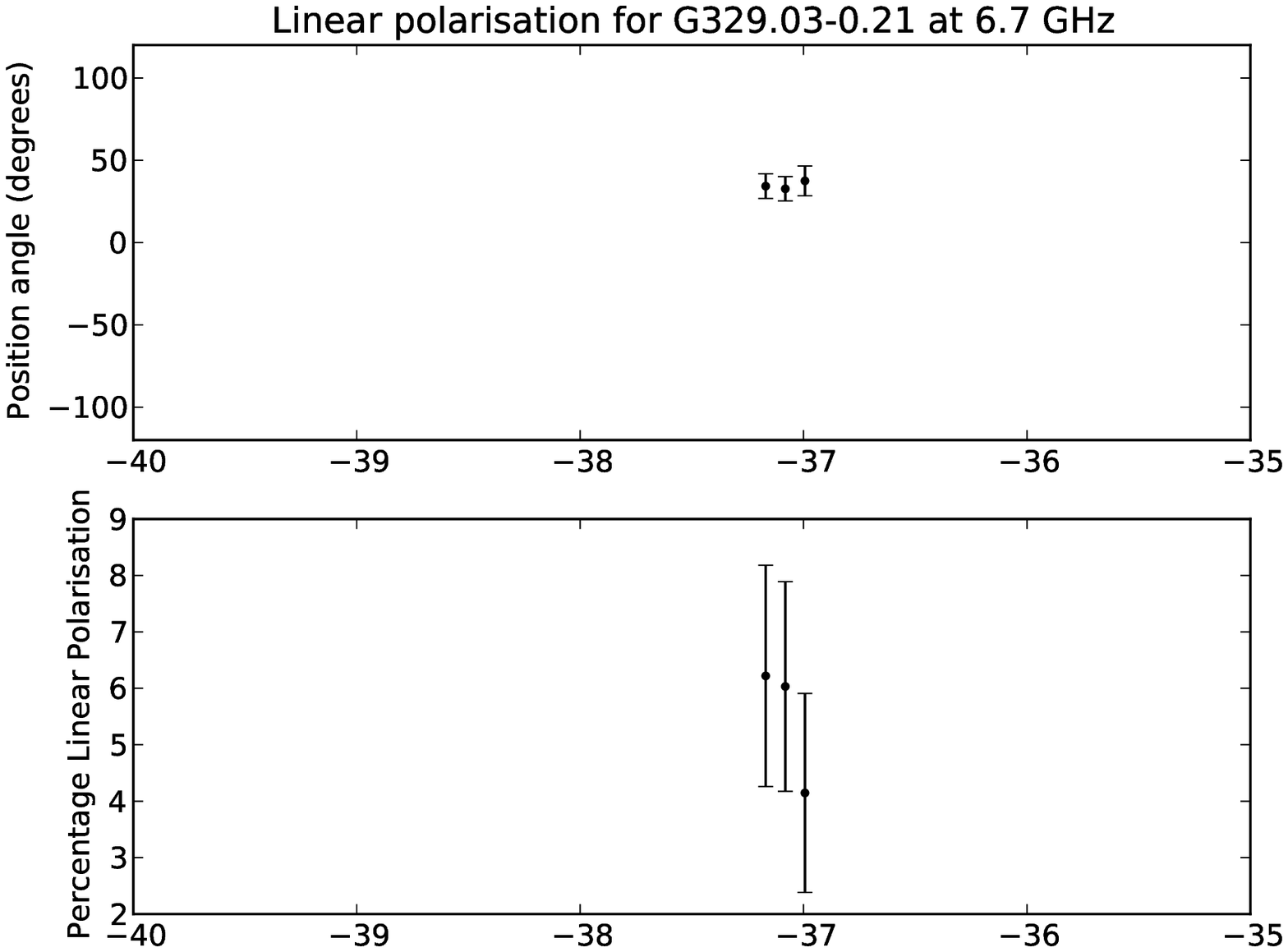}
\\
\includegraphics[scale = 0.35]{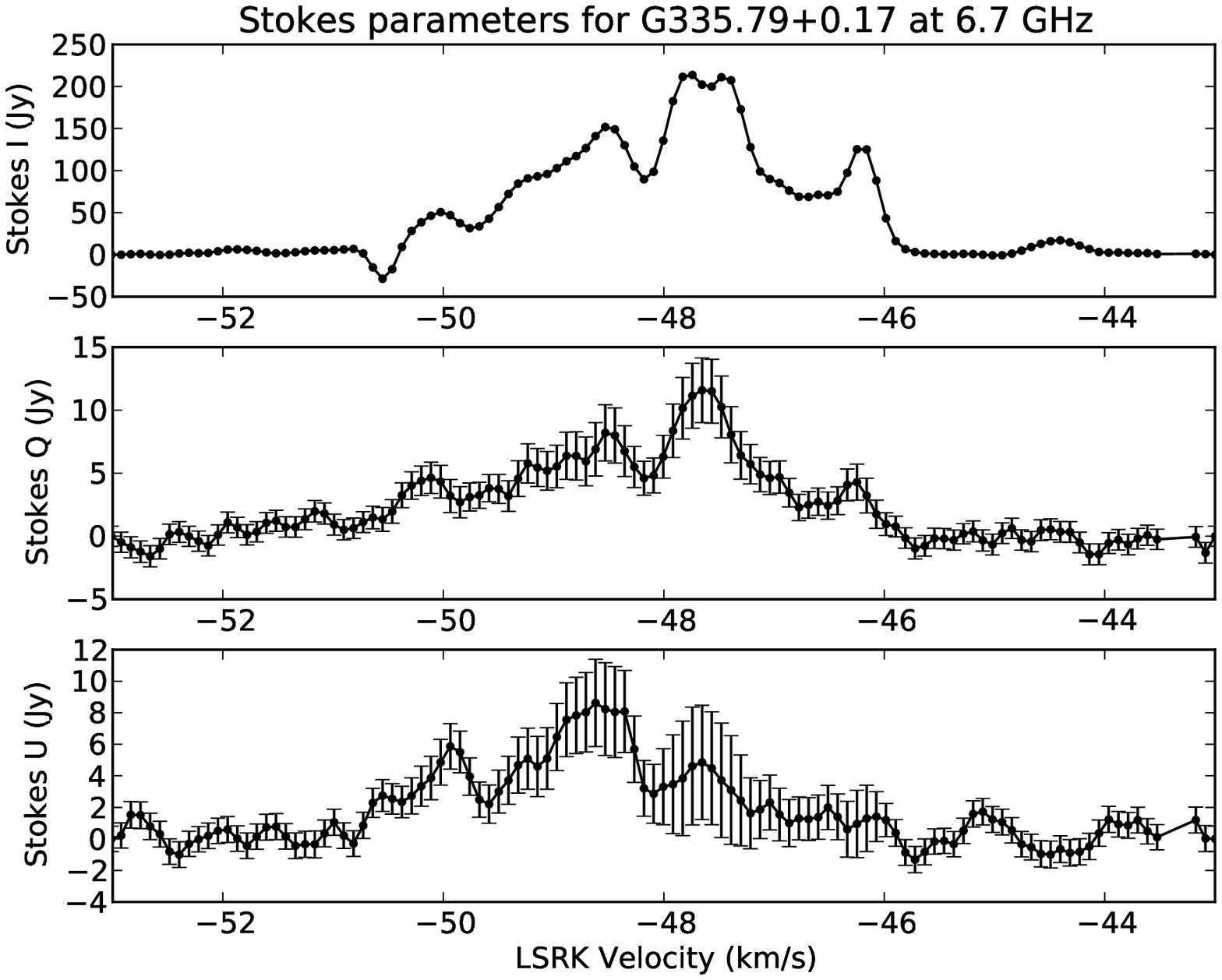}
&
\includegraphics[scale = 0.35]{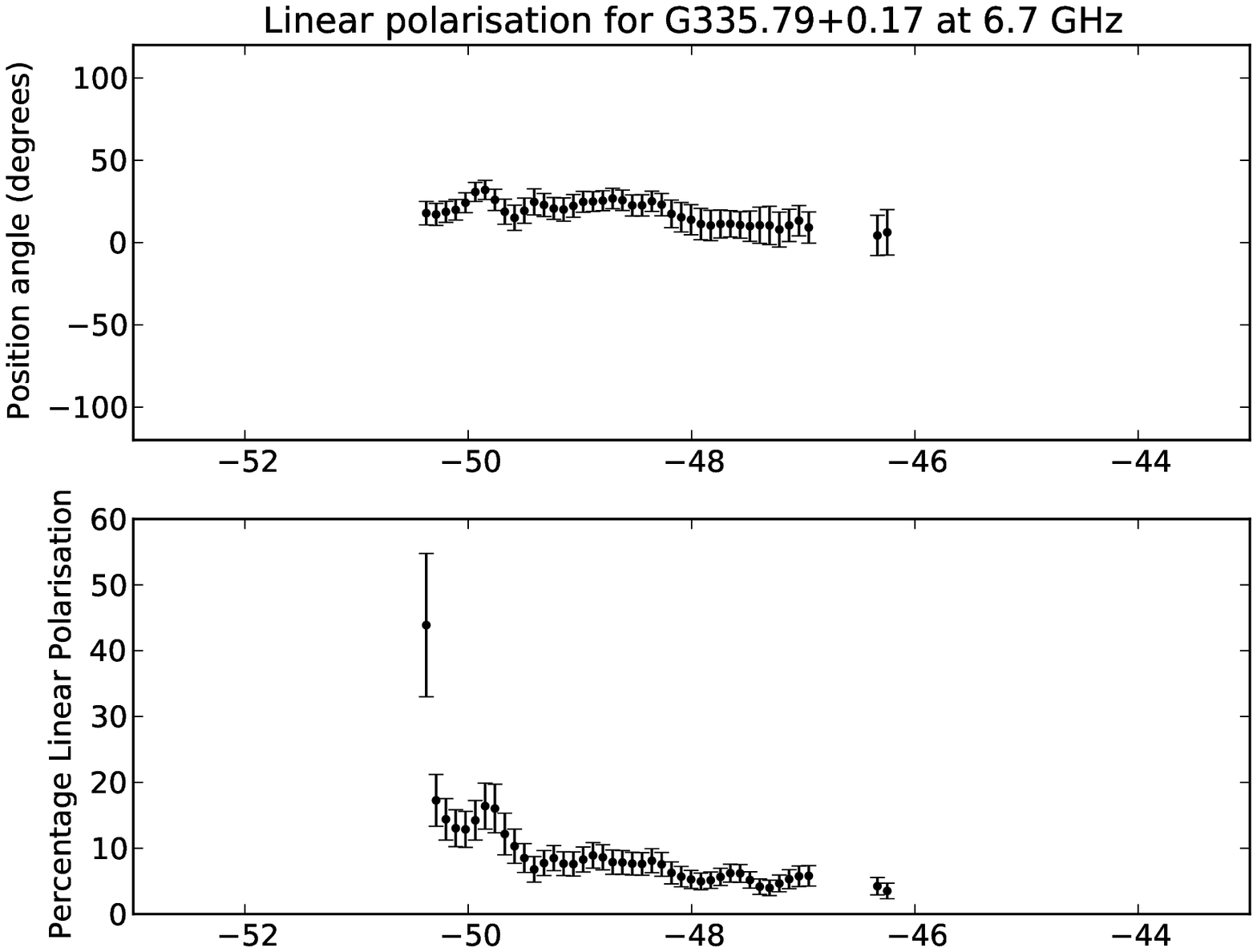}
\\
\includegraphics[scale = 0.35]{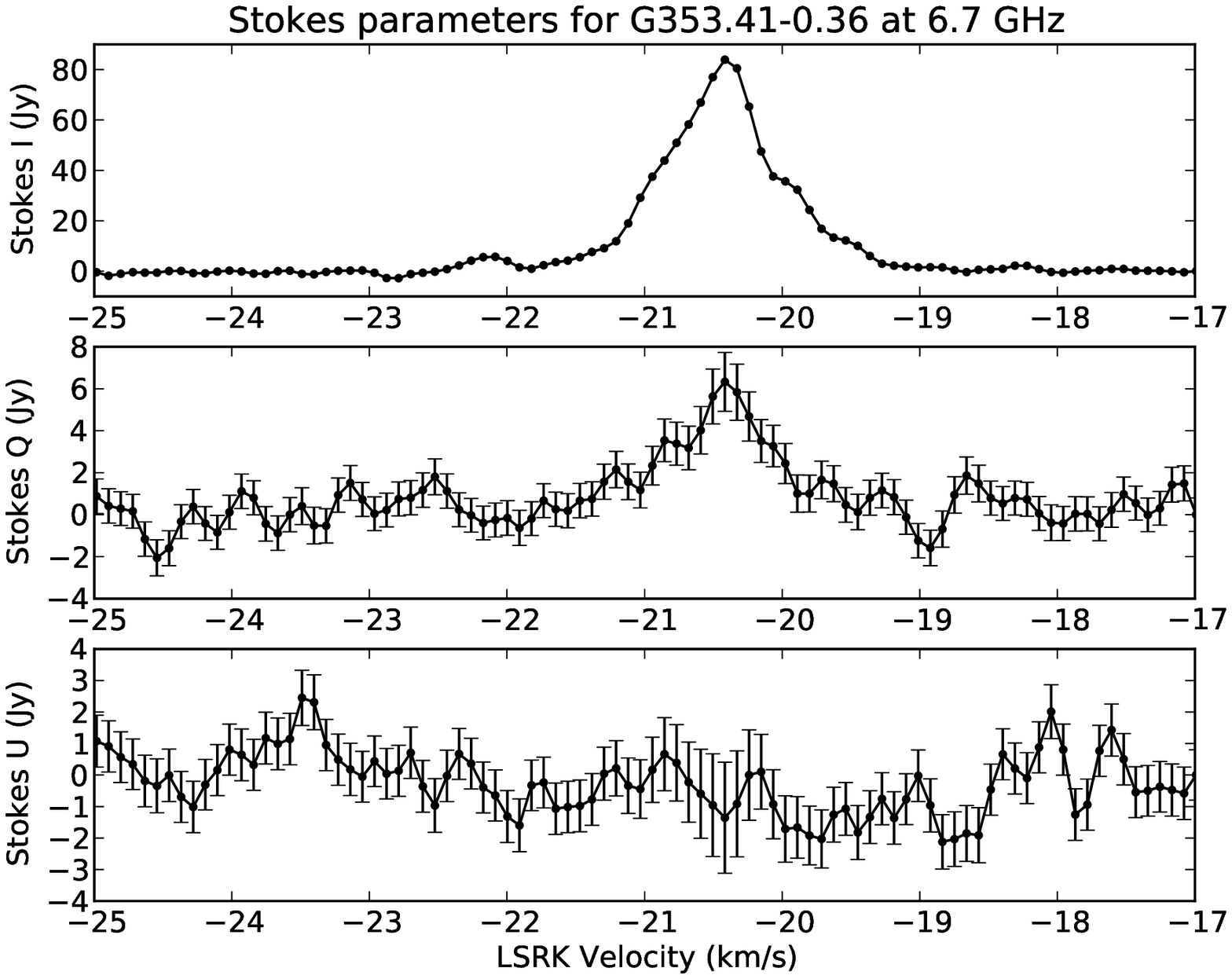}
&
\includegraphics[scale = 0.35]{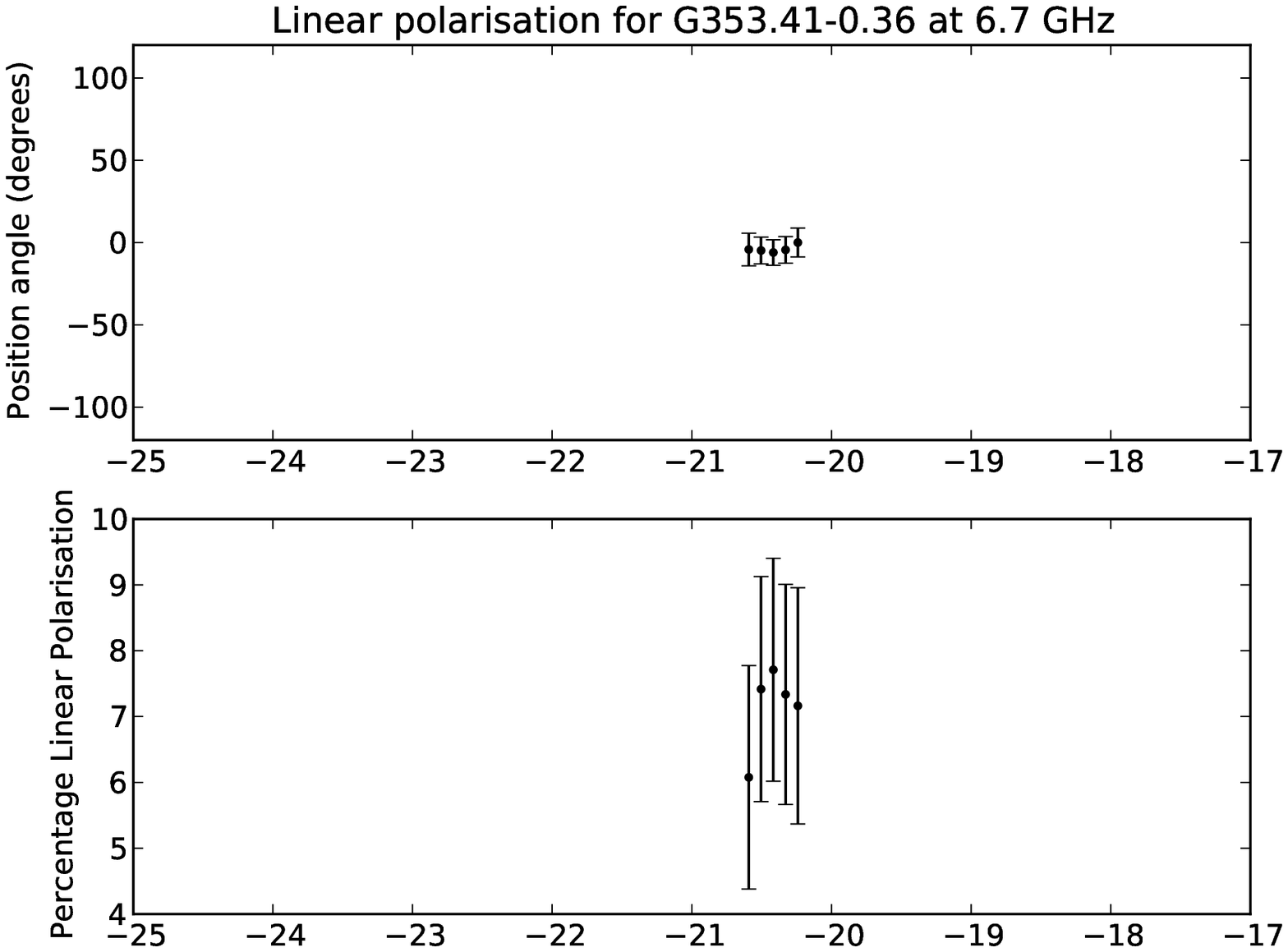}
\end{tabular}

\clearpage

\begin{minipage}{\textwidth}
\subsection*{Weaker Sources}
The level of linear polarisation in these sources was not sufficiently high to allow position angle and fractional linear polarisation plots to be produced from the observations conducted 2010 September 28-30.
\end{minipage}

\begin{tabular}{cc}
\includegraphics[scale = 0.35]{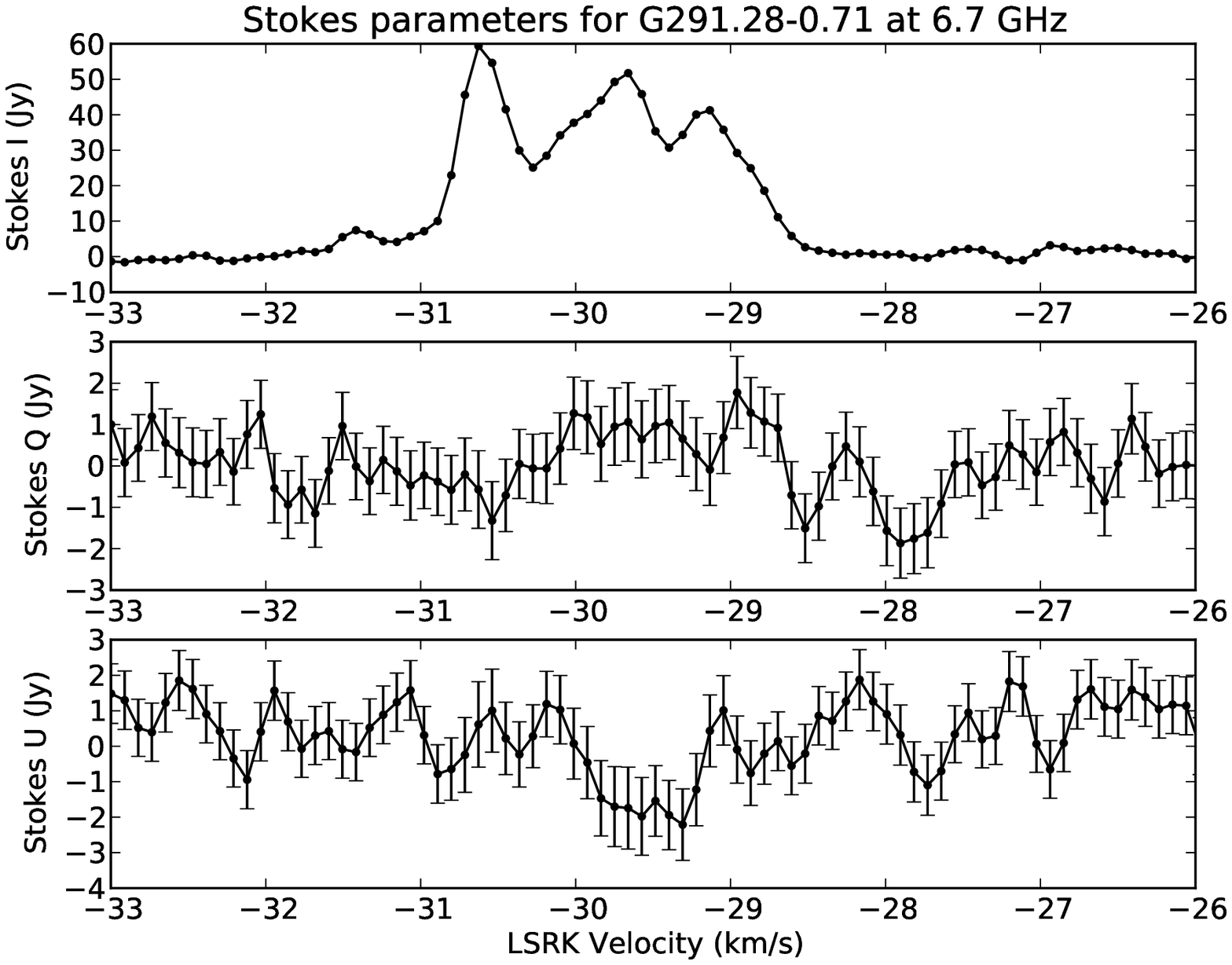}
&
\includegraphics[scale = 0.35]{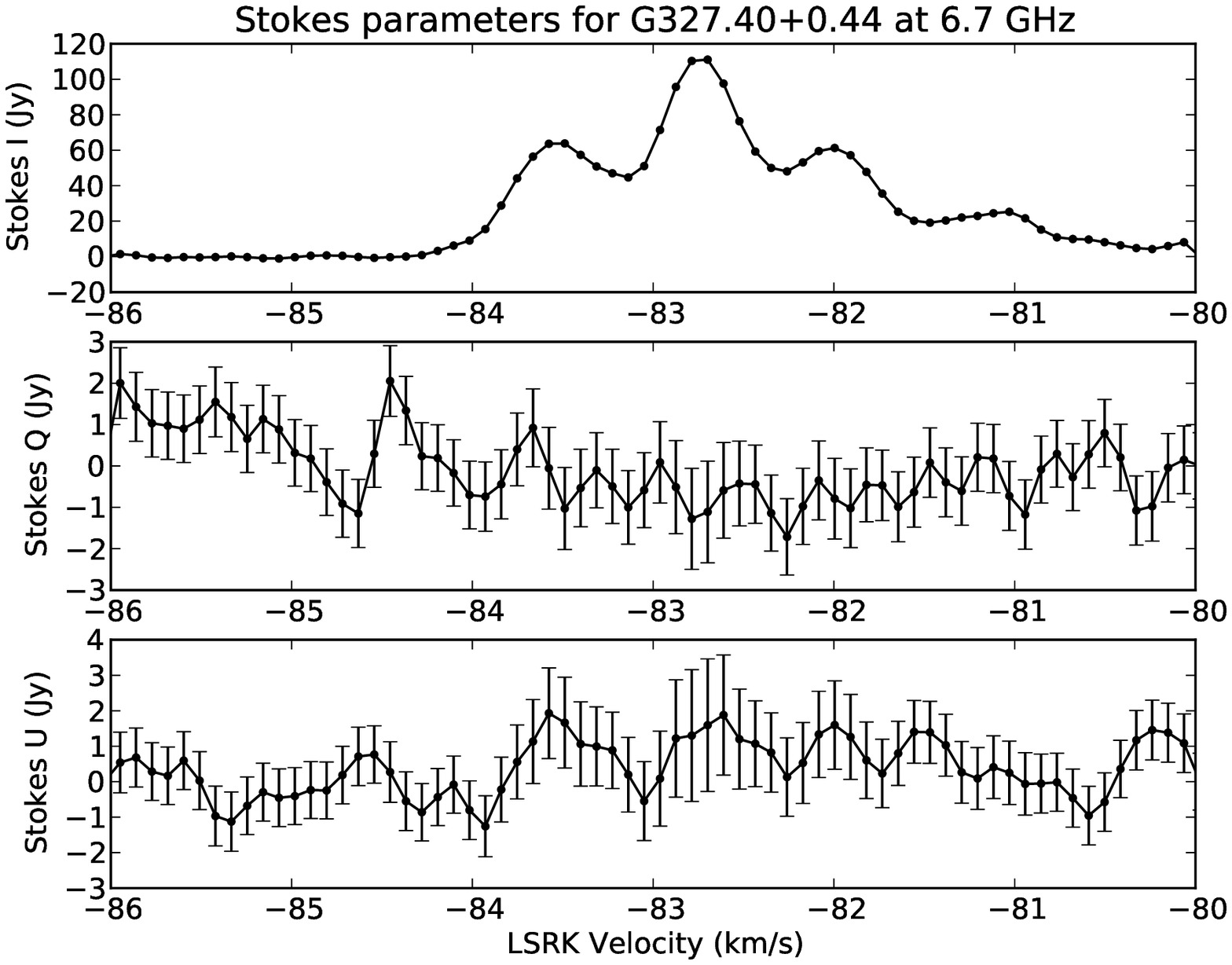}
\\
\includegraphics[scale = 0.35]{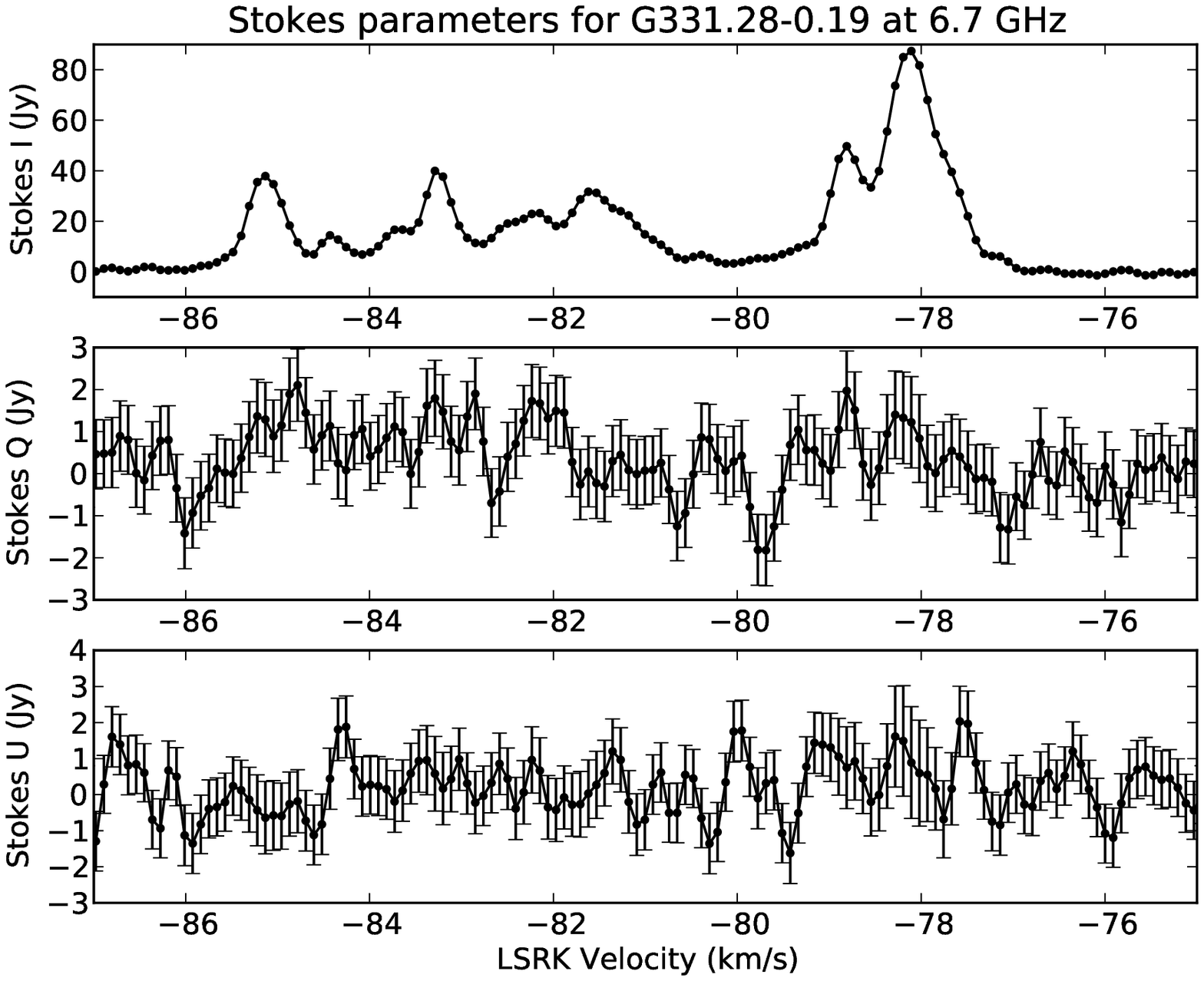}
&
\includegraphics[scale = 0.35]{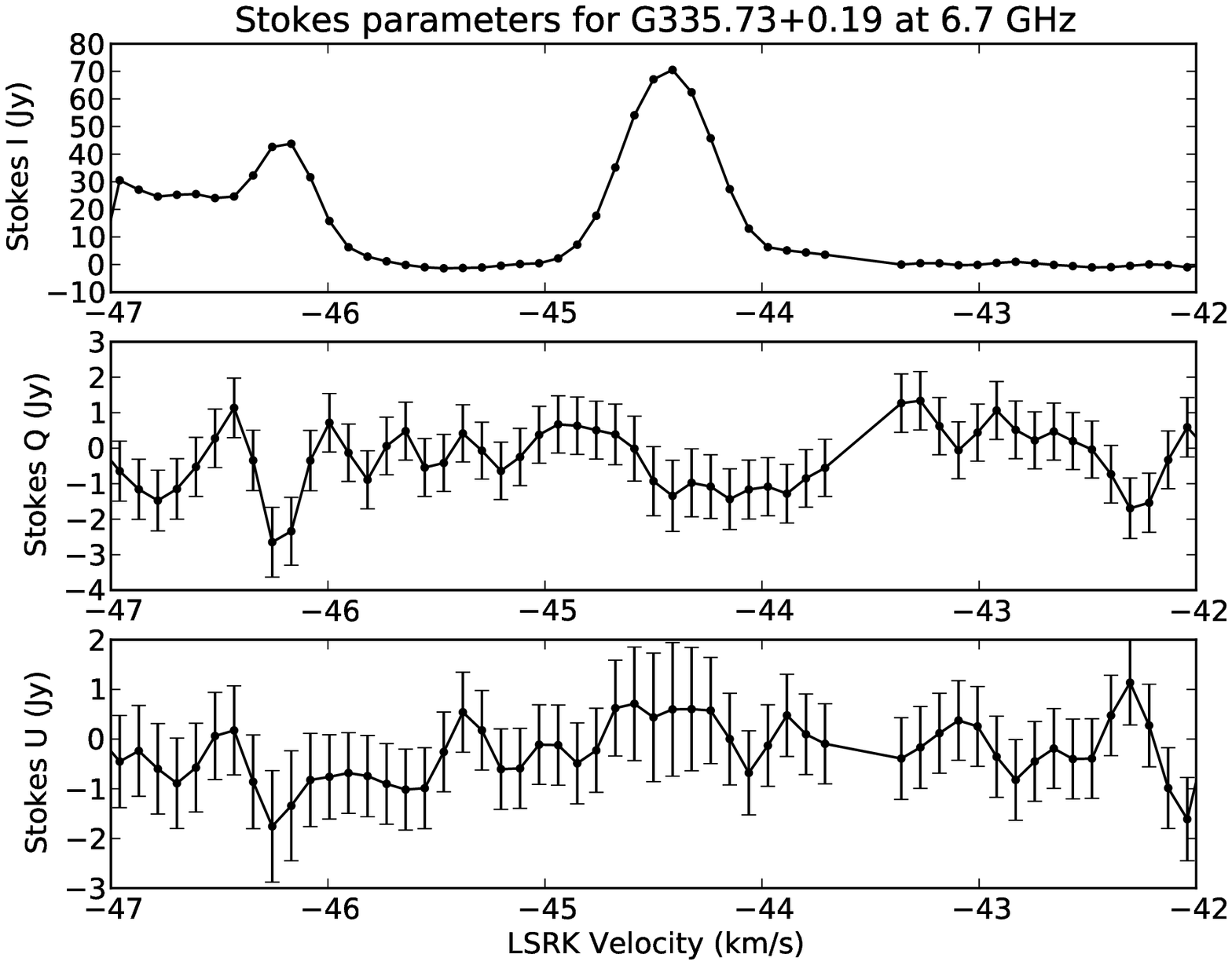}
\\
\includegraphics[scale = 0.35]{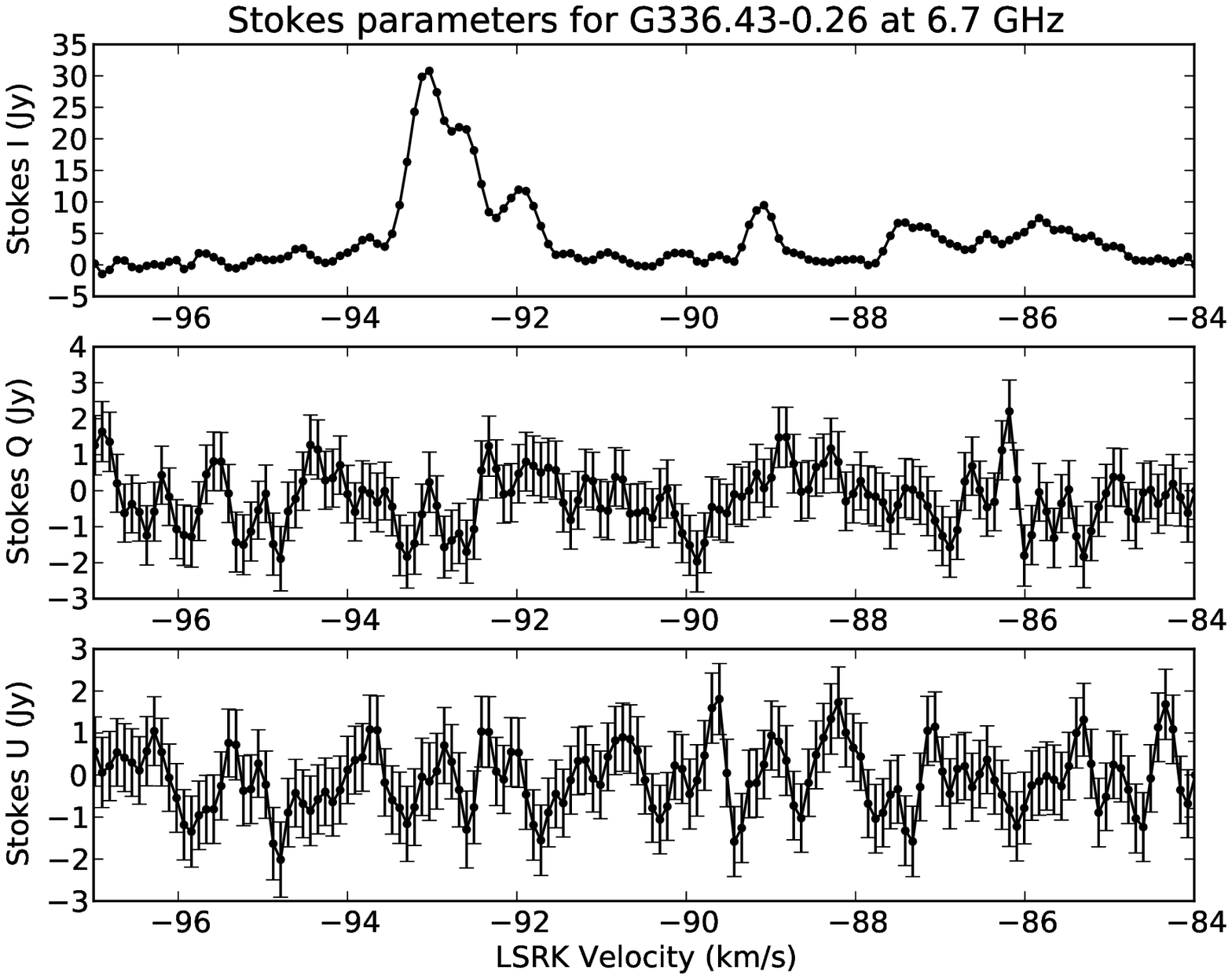}
&
\includegraphics[scale = 0.35]{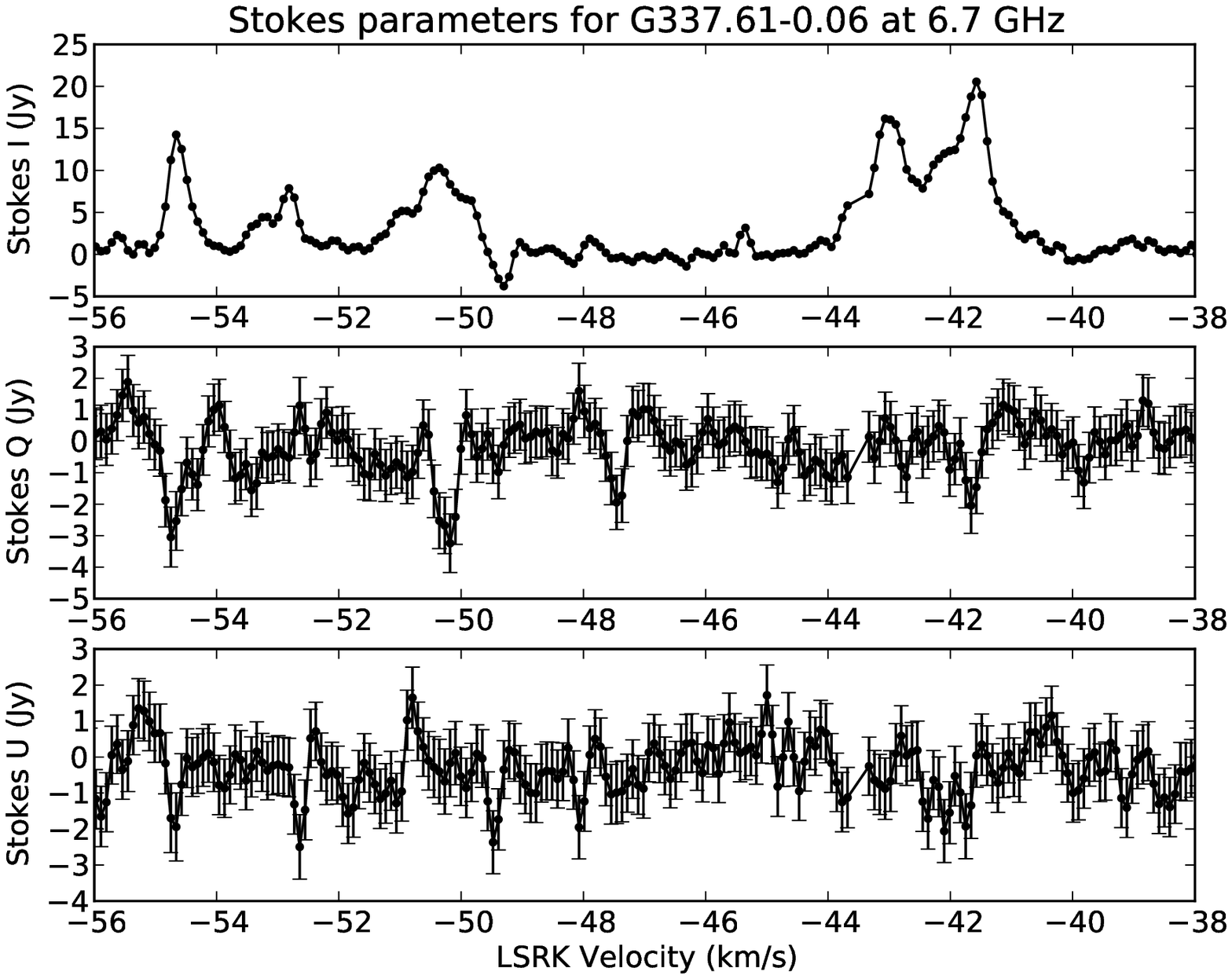}
\end{tabular}


\end{document}